\documentclass[journal, 10pt, compsoc, cspaper]{IEEEtran}

\usepackage[nocompress]{cite}
\usepackage[english]{babel}
\usepackage[T1]{fontenc}
\usepackage{xcolor}
\usepackage{amssymb}
\usepackage{amsthm}
\usepackage[skip=0pt]{caption}
\usepackage[skip=0pt]{subcaption}
\usepackage{enumitem}

\usepackage[
  placement=top,
  angle=0,
  opacity=1,
  color=black,
  scale=1.186,
  vshift=-3
]{background}
\backgroundsetup{contents={
\tiny
\begin{tabular}{c}
This article has been accepted for publication in a future issue of this
journal, but has not been fully edited. Content may change prior to final publication.
Citation information: DOI\\[1pt]
10.1109/TPDS.2020.2981891, IEEE Transactions on Parallel and Distributed Systems\\[0.808\paperheight]
This work is licensed under a Creative Commons Attribution 4.0 License. For more information, see https://creativecommons.org/licenses/by/4.0/.
\end{tabular}
}}

\usepackage{scalerel}
\newcommand{\oldstate}{{\mbox{\scaleto[1.6ex]{
  \includegraphics{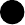}
}{1.6ex}}}}

\newcommand{\newstate}{{\mbox{\scaleto[1.6ex]{
  \includegraphics{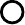}
}{1.6ex}}}}

\newcommand\orcidicon[1]{\href{https://orcid.org/#1}{{\mbox{\scalerel*{
  \includegraphics{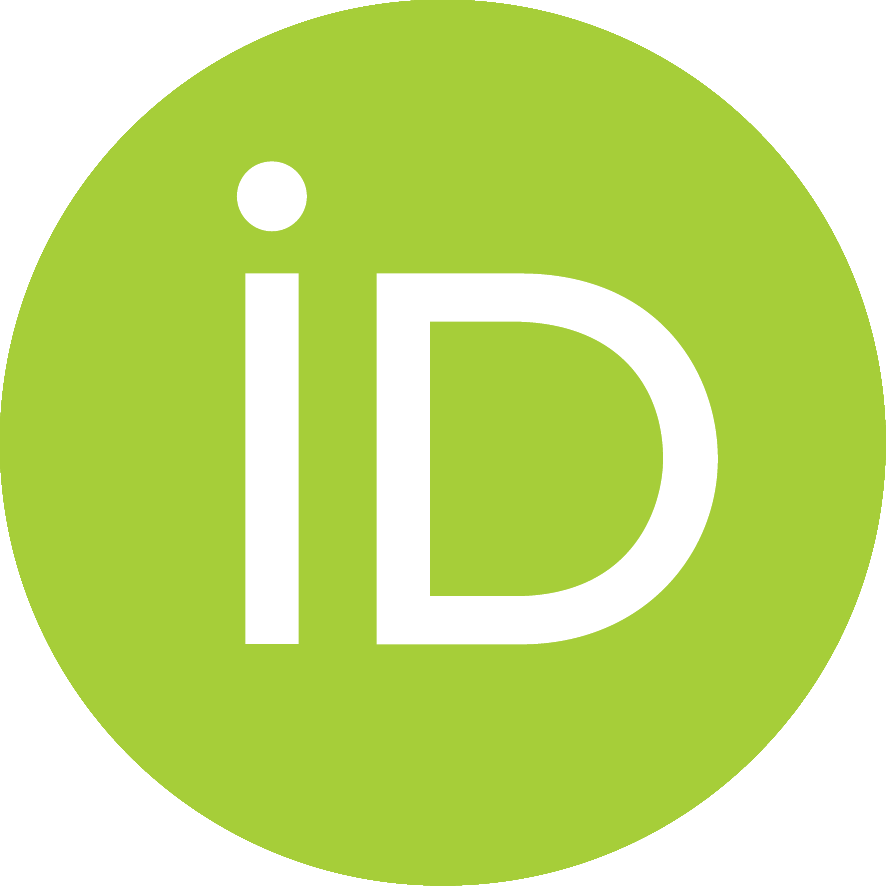}
}{|}}}}}

\usepackage[hidelinks]{hyperref} 

\let\emptyset\relax
\let\emptyset\varnothing

\newcommand{\msg}[1]{$\langle#1\rangle$}

\newcommand{\TAGprep}{\ensuremath{\mathit{PREPARE}}}
\newcommand{\TAGpaxprep}{\ensuremath{\mathit{PAXOS\_PREP}}}
\newcommand{\TAGvote}{\ensuremath{\mathit{VOTE}}}
\newcommand{\TAGvoted}{\ensuremath{\mathit{VOTED}}}
\newcommand{\TAGack}{\ensuremath{\mathit{ACK}}}
\newcommand{\TAGdone}{\ensuremath{\mathit{DONE}}}

\newcommand{\TAGlearned}{\ensuremath{\mathit{LEARNED}}}

\newcommand{\Vrack}{\ensuremath{r_\mathit{ack}}}
\newcommand{\Vrvoted}{\ensuremath{r_\mathit{voted}}}
\newcommand{\Vval}{\ensuremath{\mathit{val}}}
\newcommand{\Vreq}{\ensuremath{\mathit{req}}}

\newcommand{\Vwrite}{\ensuremath{\mathit{write}}}
\newcommand{\Vread}{\ensuremath{\mathit{read}}}

\newcommand{\Vfalse}{\ensuremath{\mathit{false}}}

\newcommand{\Pcur}{\ensuremath{_\mathit{cur}}}
\newcommand{\Pprev}{\ensuremath{_\mathit{prev}}}

\newcommand{\Fcmd}{\ensuremath{\mathit{cmd}}}

\newcommand{\Fconss}{\ensuremath{\mathit{cons_S}}}

\newcommand{\Fmaxs}{\ensuremath{\mathit{max_S}}}

\newcommand{\approach}{}
\def\approach/{RMWPaxos}

\newenvironment{proofsketch}{%
  \proof}{\endproof}
\newtheorem{prop}{Proposition}

\title{\approach/: Fault-Tolerant In-Place\\ Consensus Sequences}

\author{Jan~Skrzypczak \orcidicon{0000-0002-5423-7121},
  Florian~Schintke \orcidicon{0000-0003-4548-788X},
  Thorsten~Sch{\"u}tt \orcidicon{0000-0002-8245-5687}}%
\IEEEpubid{0000--0000/00\$00.00 \copyright 2015 IEEE}

\newcommand{\newterm}[1]{\textbf{\emph{#1}}}

\newcommand{\figref}[1]{\figurename~\ref{#1}}

\newcommand{\secref}[1]{Sect.~\ref{#1}}

\DeclareMathAlphabet{\mathcal}{OMS}{cmsy}{m}{n} 

\begin{document}

\IEEEtitleabstractindextext{%
\begin{abstract}
Building consensus sequences based on distributed, fault-tolerant consensus,
as used for replicated state machines, typically requires a separate distributed
state for every new consensus instance.  Allocating and maintaining this state
causes significant overhead. In particular, freeing the distributed, outdated states in a
fault-tolerant way is not trivial and adds further complexity and cost to the system.

In this paper, we propose an extension to the single-decree Paxos protocol
that can learn a \emph{sequence of consensus decisions}
 `in-place', i.e. with a single set of distributed states.
Our protocol does not require dynamic log structures and hence has no need
for distributed log pruning, snapshotting, compaction,
or dynamic resource allocation.

The protocol builds a fault-tolerant atomic register that supports arbitrary
read-modify-write operations. We use the concept of \emph{consistent quorums} to
detect whether the previous consensus still needs to be consolidated or
is already finished so that the next consensus value can be safely proposed.
Reading a consolidated consensus is done without state modifications
and is thereby free of concurrency control and demand for serialisation.
A proposer that is not interrupted reaches agreement on consecutive consensus decisions
within a single message round-trip per decision by preparing
the acceptors eagerly with the previous request.
\end{abstract}

}

\maketitle
\IEEEdisplaynontitleabstractindextext
\IEEEpeerreviewmaketitle

\IEEEraisesectionheading{\section{Introduction}\label{sec:introduction}}
\IEEEPARstart{S}{tate} machine replication~\cite{DBLP:journals/csur/Schneider90}
is a common technique for
implementing distributed, fault-tolerant services. Commonly,
replicated state machine (RSM) implementations are centred
around the use of a consensus protocol, as replicas
must sequentially apply the same commands in the same order to prevent divergence.

Existing consensus protocols such as Paxos~\cite{lamport2001paxos,DBLP:journals/tocs/Lamport98},
Raft~\cite{DBLP:conf/usenix/OngaroO14}, or variations
thereof~\cite{DBLP:conf/osdi/MaoJM08,DBLP:conf/dsn/JunqueiraRS11,DBLP:conf/sosp/MoraruAK13} that can
be used to build an RSM are based on the idea of a command log.
Once a replica learns one or multiple commands by consensus, it appends them
to its persistent local command log. Several practical systems~\cite{DBLP:conf/osdi/CorbettDEFFFGGHHHKKLLMMNQRRSSTWW12,
DBLP:conf/osdi/Burrows06,DBLP:conf/cidr/BakerBCFKLLLLY11} follow this general approach.

However, the implementation of such a command log incurs additional challenges
such as log truncation, snapshotting, and log recovery. In case of Paxos, these
problems have to be addressed separately on top of the consensus algorithm. This
is a challenging and error-prone task, as noted by Chandra et al.~\cite{DBLP:conf/podc/ChandraGR07}. Other
consensus solutions, e.g. Raft, consider some of these issues as part of the
core protocol while sacrificing the ability to make consensus decisions without an
elected leader. In either case, implementing consensus sequences requires extensive state
management.

A command log is worth its overhead when the commands are small
compared to the managed state. However, aggregating largely independent data
into a bigger managed state, such as multiple key-value pairs in a key-value
store, to compensate for the log overhead is counterproductive because the log
would then unnecessarily order commands targeting different keys.
Managing each key-value pair separately would be ideal, but this is
unpractical when using a log due to the implied overhead and challenges.

In this paper, we present a novel approach called \emph{Read-Modify-Write
Paxos} (\approach/) where the state
of an RSM is managed `in-place'. Instead of replicating a command log as an
intermediate step, \approach/ replicates the latest state directly. A new command is processed
by applying it to the current state and proposing the result as the next value
in a sequence of consensus decisions. Thereby, it is possible to
use a fixed set of state variables for all decisions, which avoids the
state management issues. At the same time, distributed consensus
can be used on a finer granularity than before and it becomes trivial to use an arbitrary number
of parallel consensus instances. This allows the fault-tolerant implementation of
ubiquitous primitives like counter, locks, or sets. In addition to existing
use cases like key-value stores, we believe that such fault-tolerant, fine-granular
RSM usage might become more and more relevant with the rise of byte-addressable
non-volatile memory and RDMA-capable low latency interconnects.

Before presenting \approach/, we introduce the notion of a \emph{consensus sequence register}, an
obstruction-free multi-writer, multi-reader register that performs any submitted write operation at-least-once. 
Writes are expressed in the
form of update commands applied on an opaque object. Instead of explicitly agreeing
on a sequence of commands, such register agrees on the sequence of object states that
result from the submitted update commands. By adhering to the safety properties
of consensus, reads are guaranteed to observe the latest consistent state.
Strengthening the register to apply writes exactly-once results in \approach/---a fault-tolerant general atomic read-modify-write (RMW) register.

The main contributions of this paper are:
\begin{itemize}[topsep=0pt]
\item We introduce the abstractions of a \emph{consensus sequence
  register} and strengthen it to provide an \emph{atomic RMW register}
  (\secref{sec:problem_statement}). These abstractions can be used to
  implement RSMs. If updates are idempotent the \emph{consensus
  sequence register} suffices to build an RSM.
  Otherwise, the atomic RMW register is required (\secref{sec:RSM}).
\item We provide a new implementation of a fault-tolerant atomic
  \emph{write-once} register by modifying the Paxos algorithm. In particular,
  we enhance Paxos by using the concept of \emph{consistent quorums}---a set of replies containing identical answers---to reduce contention
  in read-heavy workloads.  Once a consistent quorum is detected, the
  consensus decision is known.  This allows learning the register's value
  in two message delays and prevents concurrent reads from blocking each other (\secref{sec:cmd_sequence}).
\item By further exploiting consistent quorums, we extend the atomic
  write-once register to a multi-write register that is a \emph{consensus sequence register}. Here, a
  consistent quorum indicates the most recent consensus decision. This
  makes it possible to propose a follow-up value in-place, i.e. without
  a command log or multiple independent consensus instances (\secref{sec:cmd_sequence}). 
  If there is
  only a single writer, follow-up decisions can be made in two
  message delays (\secref{sec:optiseqwrite}) without electing a leader.

\item The \emph{consensus sequence register} applies submitted updates
  from multiple writers \emph{at-least} once, which is sufficient when
  updates manipulate the opaque object (or parts of it) in an
  idempotent way (like adding a member to a set). We show that by
  using ordered links, \emph{exactly-once} semantics can be achieved to
  build an \emph{atomic RMW register}, called \approach/
  (\secref{sec:atomic_cmd_sequence}).
\end{itemize}

\section{System Model} \label{sec:system_model}
We consider an asynchronous distributed system with processes
that communicate by message passing. Processes work at arbitrary speed,
may crash, omit messages and may recover with their internal state intact (a recovering process
is indistinguishable from one experiencing omission failures). We do
not consider Byzantine failures. A process is \emph{correct} if
it does not crash or recovers from crashes in finite time with its
(possibly outdated) state intact. We assume that every process can be identified
by its process ID (PID).

In the first part of this paper, processes send messages to each other via direct unreliable
communication links. Links may lose or delay messages indefinitely or
deliver them out-of-order. While a fair-loss property~\cite{DBLP:books/daglib/0025983}
is desirable to support progress, it is not formally necessary. In \secref{sec:atomic_cmd_sequence}, we strengthen this and require reliable
in-order message delivery. Such reliable links can easily be constructed on
top of unreliable fair-loss links~\cite{DBLP:books/daglib/0025983}. In practice,
TCP is often used as reliable communication protocol~\cite{DBLP:conf/dsn/JunqueiraRS11}.

\section{The Consensus Problem} \label{sec:paxos_background}
The consensus problem describes the agreement of several processes
on a common value in a distributed system. We differentiate between
\newterm{proposer} processes that propose values and \newterm{learner}
processes that must agree on a single proposed value. In practice, a process can also
implement both roles.  A correct solution to the consensus
problem must satisfy the following \newterm{safety}
properties~\cite{lamport2005generalized}:

\begin{description}
    \item[C-Nontriviality.] Any learned value must have been proposed.
    \item[C-Stability.] A learner can learn at most one value.
    \item[C-Consistency.] Two different learners cannot learn different values.
\end{description}

In addition to safety, the \newterm{liveness} property requires that some value is eventually
learned if a sufficient number of processes are correct. However, guaranteeing
liveness while satisfying the safety properties of consensus is impossible in
an asynchronous system with one faulty process~\cite{DBLP:journals/jacm/FischerLP85}.

\section{Problem Statement} \label{sec:problem_statement}
We define a fault-tolerant register that is replicated on $N$
processes and tolerates the crashes of a minority of replicas. The
register holds a value $v$. Its initial value is $v=\bot$.
Any number of clients can read or modify $v$ by submitting
\emph{commands} to any replica. The primary motivation of our work is to provide
a register abstraction that allows the implementation of a replicated
state machine. For that, we start with a simpler abstraction, which we then extend.

\begin{description}[listparindent=\parindent, itemindent=\parindent, leftmargin=0cm]
    \item[Write-Once Atomic Register.]
        Commands submitted to the register either write a value or read its
        current value. Read commands return either $\bot$ or a value $v_w$ that
        has been submitted by some write.
        The register is linearisable~\cite{DBLP:journals/toplas/HerlihyW90}, i.e.
        all commands appear to take effect instantaneously at some time between
        their submission and the corresponding completion response from the register. Thus,
        once a read returns $v_w$, then all subsequent reads must return $v_w$ as well. However,
        an arbitrary number of reads is allowed to return $\bot$ beforehand if no value was written yet. This is achieved by satisfying the
        safety properties stated in \secref{sec:paxos_background}.
    \item[Consensus Sequence Register.]
        We extend the write-once atomic register by allowing multiple clients to submit
        \emph{update} commands that change the register's value.
        We say that a value $v$ is the result of
        \emph{update sequence} $s(v) = u_1, \ldots, u_n$, iff $v$ equals $u_n \circ \dots \circ u_1$
        applied on $\bot$ ($\circ$~being function composition).
        The register ensures that reads return values with growing update sequences.
        For that, we extend the safety properties of consensus for
        consensus sequences.
        \begin{description}[leftmargin=\leftmargini]
            \item[CS-Nontriviality.] Any read value is the result of applying a sequence of
                submitted updates.
            \item[CS-Stability.] For any two subsequent reads returning values
              $v_1$ and $v_2$, $s(v_1)$ is a prefix of $s(v_2)$.
            \item[CS-Consistency.] For any two reads
              (including concurrent ones) returning values $v_1$ and $v_2$,
              $s(v_1)$ is a prefix of $s(v_2)$ or vice versa.
        \end{description}
        The prefix relation on update sequences is reflexive. Every
        update sequence is its own prefix. Update sequences are merely
        a tool to argue about the register's properties. The actual register implementation
        does not explicitly store them. It simply keeps the value resulting
        from the latest update.

        For updates, we also require the following properties:
        \begin{description}[leftmargin=\leftmargini]
            \item[CS-Update-Visibility.] Any completed update is
              included at least once in the update sequence of all
              values returned by subsequent reads.
            \item[CS-Update-Stability.] For any two subsequent updates
            $u_1$ and $u_2$, $u_1$ appears before $u_2$ in the update sequence of
            any returned value that includes both $u_1$ and $u_2$.
        \end{description}

    \item[Atomic Read-Modify-Write Register.]
        To satisfy linearisability, we strengthen CS-Update-Visibility by requiring
        that every completed update is included \emph{exactly-once} in the update sequence
        of all values returned by subsequent reads. This results in a general atomic
        read-modify-write (RMW) register~\cite{DBLP:books/mk/Lynch96}.
        Unlike specialised RMW registers that
        can perform a single type of RMW operation like test-and-set or fetch-and-add,
        this register can atomically execute arbitrary computations on its previous value.
\end{description}

As liveness is impossible in our system model, wait-freedom~\cite{DBLP:journals/toplas/Herlihy91}
cannot be provided. However, we require obstruction-freedomness~\cite{DBLP:conf/icdcs/HerlihyLM03}
for a valid implementation of the registers. If wait-freedom is still required,
an obstruction-free implementation can be extended by a leader oracle 
assuming a $\Diamond\mathcal{W}$ failure detector~\cite{DBLP:journals/jacm/ChandraHT96}.

\section{In-Place Consensus Sequence} \label{sec:implementation}
In this section, we present our protocols that
satisfy the properties of the register abstractions introduced
in \secref{sec:problem_statement}. The write-once atomic register makes use of the
principles of Paxos consensus~\cite{DBLP:journals/tocs/Lamport98,lamport2001paxos}
and adopts the concept of \emph{consistent
  quorums}~\cite{consistent-quorum}.
These concepts are then extended for the more powerful abstractions
to allow a sequence of multiple consensus decisions `in-place', i.e.
on the same set of state variables by overwriting the previous consensus.
A more detailed, albeit more informal description of a previous version
is given by Skrzypczak~\cite{Skrzypczak2017}. We discuss how to build an
RSM with our register in \secref{sec:RSM}.

\subsection{Paxos Overview}
Our approach is derived from the Paxos
protocol. In addition to proposers and
learners, Paxos introduces the role of \newterm{acceptor} processes that coordinate
concurrent proposals by \newterm{voting} on them. If a sufficient number of
acceptors have voted for the same proposal, the proposal's value can be learned
by a learner. Such a set of acceptors is called a \newterm{quorum}. A
proposal is \newterm{chosen} if it has acquired a quorum of votes. The value of
a chosen proposal is a chosen value. The
size of quorums depends on the application and Paxos variant in
use~\cite{DBLP:journals/dc/Lamport06,DBLP:conf/opodis/HowardMS16,DBLP:conf/sosp/MoraruAK13}.
However, it is generally required that any two quorums have a
non-empty intersection to prevent two disjoint quorums that voted for different
values (as this would allow two learners to learn different values).

For Paxos to learn a value, a quorum of acceptors, a learner, and
the proposer that has proposed the value, must be correct during the execution
of the protocol. For simplicity, we consider any majority of acceptors to be
a quorum. Thus, a system with $2F + 1$ acceptors can tolerate at
most $F$ acceptor failures.

If enough processes are correct, then Paxos is obstruction-free~\cite{DBLP:conf/icdcs/HerlihyLM03},
i.e. an isolated proposer without concurrent access succeeds in a finite number
of steps. However, concurrent proposals can invalidate each other repeatedly,
thereby preventing learners from learning any value. This scenario is known
as \newterm{duelling proposers}.

\subsection{Consistent Quorums}\label{sec:consistent-quorum}
\begin{figure}[!t]
    \centering
    \scalebox{0.9}{\includegraphics{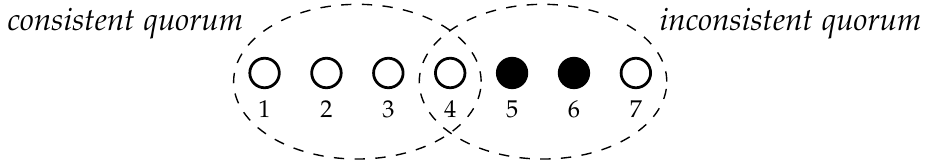}}
    \vspace*{2mm}
    \caption{Consistent/inconsistent quorum with 7 acceptors. A quorum
    view $Q$ for a system using $n$ acceptors consists of
    $|Q| = \lfloor\frac{n}{2}\rfloor+1$ elements (here 4).}
    \label{fig:cons_quorum}
\end{figure}

Similar to Paxos, our approach structures the communication between
proposers and acceptors into phases. In each phase, a proposer sends a
message to all acceptors and waits for a minimal quorum of replies.
The seen quorum is
\newterm{consistent} if the indicated state by the acceptors in the
quorum is identical, otherwise, it is \newterm{inconsistent}
(see \figref{fig:cons_quorum}). Not waiting for more replies than
necessary ensures tolerating a minority of failed acceptors without
delaying progress.

If a proposer $p$ observes an inconsistent quorum, it cannot infer
which of the seen values is or will be chosen and learned.  For example, if $p$
receives the quorum depicted in the right part of \figref{fig:cons_quorum}, it cannot
decide if \oldstate\ or \newstate\  exists in a majority since it has no
information about the state of acceptors $1$--$3$.
In contrast, it is trivial for $p$ to deduce the chosen
value with a consistent quorum (\figref{fig:cons_quorum} left).
Existing Paxos variants do not distinguish consistent or inconsistent
quorums. As we will see, detecting a consistent quorum allows the
proposer to terminate the protocol early in the single-decree case.
Furthermore, the consistent state can be used as the basis for follow-up proposals
if multiple consensus decisions are needed in sequence.

\subsection{Paxos Write-Once Atomic Register} \label{sec:write-once}
In the following, we present our modifications to the original
single-decree Paxos protocol for implementing a write-once atomic
register. Its pseudocode is depicted in Algorithm~1.
We note that no separate learner role exists, as each proposer also
implements the functionality of a learner in our implementation.
To make the algorithm easier to understand,
we provide an execution example in Figure~\ref{fig:flow}.
We discuss differences to Paxos in \secref{sec:comparison-paxos}. Before
proceeding to the algorithm description, we first cover some general concepts and conventions.

\begin{description}[listparindent=\parindent, itemindent=\parindent, leftmargin=0cm]

\item[Rounds.]
Concurrent proposals are ordered by so-called \newterm{rounds}
(analogue to `proposals numbered $n$' in~\cite{lamport2001paxos} and
`ballot numbers' in~\cite{DBLP:journals/tocs/Lamport98}).  A round is a tuple $(n, id)$,
where $n$ is a non-negative integer and $id$ some globally unique identifier.
Rounds are partially ordered. $r_1 < r_2$ iff $r_1.n < r_2.n$.
Furthermore, $r_1 = r_2$ iff $r_1.n = r_2.n \land r_1.id =
r_2.id$. Newer proposals are indicated by higher rounds. Rounds with
the same $n$ but different $id$ cannot be ordered.

\item[Acceptor State.]
Acceptors act as the distributed, fault-tolerant storage. Each acceptor
manages three values (cf. Algorithm~1, line 24): (1) the
highest round \Vrack\ it has acknowledged,
(2) the last value \Vval\ it has voted for, and (3) round \Vrvoted\
in which the proposal including the value was proposed in. By acknowledging a round,
acceptors promise not to vote for lower-numbered proposals in the future.

\begin{figure*}[t]
    \includegraphics[width=\textwidth]{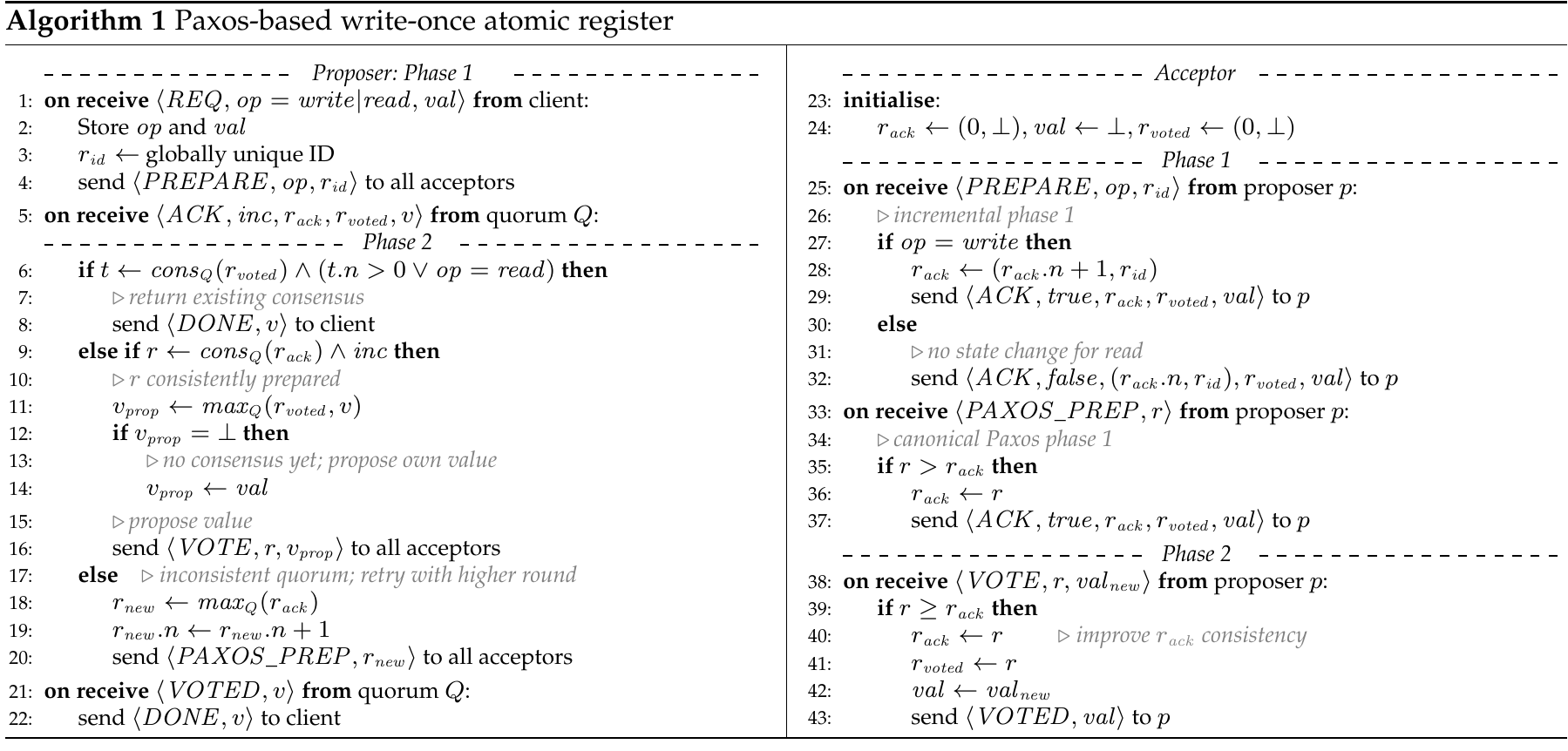}
    \vspace{-1.3em}
\end{figure*}

\item[Pseudocode Conventions.]
For brevity's sake, we use the following conventions when handling
sets of reply messages: Let a process receive a set of reply messages
$S$. Each message in $S$ is an $n$-tuple denoted as \msg{t, e_1,\dots,e_{n-1}}.
We make use of pattern matching techniques commonly found in functional programming. The
type $t$ of the message is matched to ensure it has the correct format. Its
payload is stored in tuple elements $e_1$ to $e_{n-1}$. Since messages in $S$ may hold different
values in the same tuple element, we define the following functions:
$\Fconss(e_i)$ returns the value of $e_i$ if it is equal for all messages in $S$,
or \Vfalse\ otherwise;
$\Fmaxs(e_i)$ returns the largest value of $e_i$;
$\Fmaxs(e_i, e_j)$ returns the value of $e_j$ from the message with the largest value of $e_i$.

We furthermore assume that processes can keep track of multiple concurrent requests and
know to which outstanding request a received reply belongs.
\end{description}

\subsubsection{Protocol Description}
The protocol has two phases. In the
first phase, a proposer checks for concurrently proposed values and prepares
acceptors to deny outdated proposals. In the second phase, a proposer
proposes either its own or a value seen in the first phase. To eventually
learn a value, both phases must be passed without interruption by other proposers.

The protocol begins with proposer $p$ receiving a request from a client (line~1).
The request is either a \Vwrite\ that tries to set the register to a value \Vval,
or a \Vread\ that returns the register's current value (here, $\Vval = \bot$).
The request of the client is handled asynchronously. The client will be notified by
a \TAGdone\ message once the request has been processed.

Proposer $p$ starts the first phase by choosing a round ID and
sending it in a \TAGprep\ message along with the request type to all acceptors (line~2--4).
Any acceptor $\mathcal{A}$ that receives a \Vwrite\ request from $p$ acknowledges this
by incrementing \Vrack\ and updating its ID. Thereby, $\mathcal{A}$
promises $p$ to not vote for any lower-numbered proposals in the future (line~28--29).
If $\mathcal{A}$ received a \Vread\ request, then it does not increment \Vrack\
as $p$ does not intend to modify the register's value by submitting a proposal.
Letting the state untouched when processing reads reduces their interference with
other ongoing requests and is not part of canonical Paxos.

After processing the request, $\mathcal{A}$ replies with its current state and
indicates if its \Vrack\ round was incremented (lines~29,~32).
The second phase begins as soon as $p$ has received replies from a quorum $Q$ of
acceptors. Depending on the replies, $p$ proceeds in one of the following ways:

(1) If all acceptors in $Q$ have voted for the same proposal (same \Vrvoted), then $p$
knows that consensus was already reached and that the proposal's value is chosen.
Thereby, $p$ has learned the register's value and returns it to the client.
Similarly, $p$ can be certain that consensus was not reached if no acceptor in
$Q$ has voted for any proposal yet. Thus, it can return an empty value
if it is processing a \Vread\ (line~6--8).

(2) If all acceptors in $Q$ incremented their rounds and responded with a consistent
\Vrack\ round, then $p$ can propose a value. If at least one of the acceptors has voted
for a past proposal, $p$ receives an inconsistent quorum as shown in \figref{fig:cons_quorum}.
It cannot decide if the proposal's value is already established or not. In order
to not violate safety, $p$ must propose the value seen in the
highest round. If no acceptor has voted for any proposal yet, $p$ can choose its
own value. The proposal is sent in a \TAGvote\ message to all acceptors using the
acknowledged round \Vrack\ (line~9--16).

\begin{figure*}[t]
    \centering
    \resizebox{.88\textwidth}{!}{%
      \includegraphics{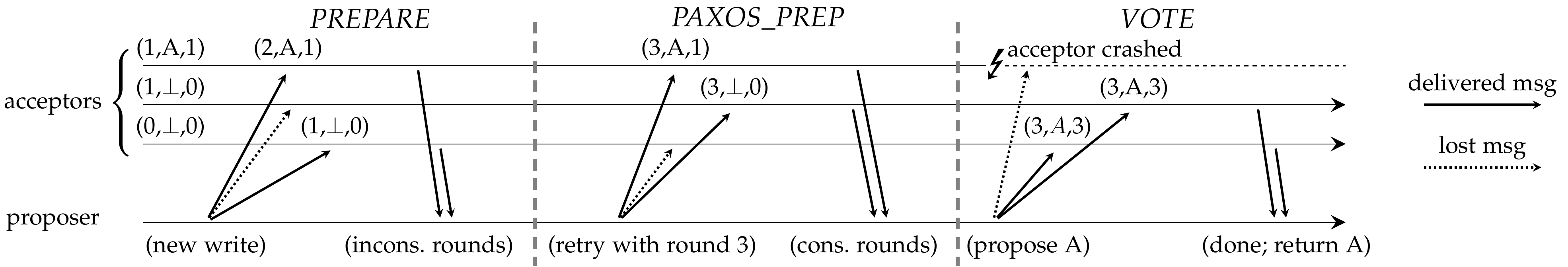}
    }
    \captionsetup{justification=centering,margin=1cm}
    \caption{Example message exchange of a write, starting with inconsistent acceptor states. Time moves from left to right.
     Acceptor states are shown as ($r_{ack}$, $\textit{val}$, $r_{voted}$). Round IDs are omitted for simplicity.}
     \label{fig:flow}
     \vspace{-1em}
\end{figure*}

(3) In all other cases, $p$ has to retry the first phase. This happens if
acceptors are currently in an inconsistent state, e.g. because of an ongoing
proposal, lost messages, or a crashed proposer. As $p$ has already knowledge 
about the current state of the acceptors, it can choose an explicit round
number that is higher than all rounds observed so far, which is then included
in \TAGpaxprep\ messages (line~17--20).
An example of this is depicted in \figref{fig:flow}.

Each acceptor that has received a proposal by $p$ (case~(2)), votes
for the proposal if they have not given a promise for a higher round during
a (concurrent) phase 1 and notifies $p$ of its vote (line~40--43).
Otherwise, the acceptor ignores the proposal or may optionally
notify $p$ that its proposal is outdated (not shown).
Once $p$ has received a quorum of positive replies, it knows that
its proposed value is chosen and notifies the client on the established consensus
(line~22). This concludes the protocol.

\subsubsection{Comparison to Paxos}\label{sec:comparison-paxos}
Our write-once atomic register is based on the same mechanism for
safety as Paxos, but differs from the canonical single-decree
Paxos~\cite{lamport2001paxos} in several aspects:

\begin{description}[listparindent=\parindent, itemindent=\parindent, leftmargin=0cm]

\item[Consistent Quorums.] In canonical Paxos, all proposals must
  complete both phases of the algorithm even if a value was already
  chosen. This effectively serialises concurrent reads and causes
  unnecessary state changes in acceptors (their round numbers).  Our
  protocol, instead, terminates early and returns the result after the
  first phase, when a proposer observes a consistent quorum.  This
  prevents (1) state modifications by reads, (2) allows termination in two
  message delays and (3) prevents live-locks caused by duelling proposers
  once all correct acceptors have agreed on a proposal. This is
  possible because once a proposal with value $v$ is chosen, any
  proposal made in a higher round will contain $v$ (see
  \secref{sec:proof}). As the value of the register cannot
  change any more, it is needless to execute the second phase.

\item[Distinguishing between reads and writes.]  In canonical Paxos,
  to read the state of a consensus it is necessary to propose a
  value for consensus when no proposal was seen yet, i.e. actually
  performing a write, which is unintended. For a read, a client can either
  (1) initiate the protocol as a proposer and---in accordance to the
  protocol---has to propose a (dummy) value itself when no value was
  chosen yet or (2) it can ask a learner. However, a learner that has
  not learned a value also has to propose a (dummy) value to
  ensure its answer is up-to-date. As this dummy value might be
  written, the read semantic is violated.
  Drawing from the concept of consistent quorums, we support reads
  without the risk to  change the register's value and are also able to
  reliably recognise an empty register. A read acts like a write 
  only when an ongoing, partially accepted proposal is seen that may need help
  to fully establish.
  However, no value will be proposed that was not already proposed by a write.

\item[Incremental round number negotiation.]
  Proposers have to choose a high enough round number for their
  proposal to succeed. Canonically, a proposer chooses the round number itself.  If
  it is too low, the proposer's attempt fails and it has to try
  again with a higher round. This works
  well when a leader makes the proposals, as it knows the
  previous used round number. Without a leader, however,  the first guess of
  a proposer is likely to fail, costing a round trip even without concurrent access.
  Instead, we let the acceptors increment their round on an initial
  round-less attempt and retrieve the `assigned' round from the
  replies when they form a consistent quorum. Otherwise, we calculate a
  higher round number from the replies and retry like in Paxos.

  Using incremental rounds is optional. If a proposer can determine a
  round number that likely succeeds, it can also start with that
  without violating the protocol's safety.

\item[Single learner per request.]
In canonical Paxos, acceptors send their \TAGvoted\ messages to a set of learner
processes, which learn the value once they have received a quorum of votes for a
proposal. Therefore, the number of messages sent is the product of the number of
acceptors and the number of learners. In our approach, the proposer that has received
a request acts as its sole learner. Thus, every acceptor sends only a
single \TAGvoted\
message. 
\end{description}

\subsubsection{Sketched Proof of Safety} \label{sec:proof}
In this section, we provide a proof sketch for our Paxos atomic write-once register.
We show that the safety requirements of \secref{sec:paxos_background}, as well as
linearisability are satisfied.
Since our protocol has a close resemblance to canonical Paxos, we can use analogue
arguments and invariants as described by Lamport~\cite{lamport2001paxos} to prove safety.

\begin{prop} \label{prop:1}
    If a proposal $p$ was learned in round $r$, then there exists a
    quorum of acceptors $Q$ such that any acceptor in $Q$ has given a
    vote for $p$ (i.e. the proposal must have been chosen).
\end{prop}
\begin{proofsketch}
For any two acceptors $a_1$, $a_2$, which have voted for proposal $p_1$ and $p_2$
respectively in the same round $r$, it holds that $p_1=p_2$ because rounds are
uniquely identified by their ID. To learn a value, a proposer must either (a)
receive a consistent quorum of \Vrvoted\ rounds from acceptors at the beginning
of phase 2, or (b) receive a quorum of \TAGvoted\ messages. For (a) to be possible, a
quorum with $r=\Vrvoted$ must exist. For (b), a quorum of acceptors
must have voted for the proposer's proposal. Since a proposal is issued for
a specific round, all replying acceptors have voted for a proposal in the
same round.
\end{proofsketch}

C-Nontriviality is trivial to proof using proposition~\ref{prop:1}
since acceptors can only vote for any value
that was previously proposed by a proposer. C-Stability and C-Consistency hold by
satisfying the following invariant:
\begin{prop} \label{prop:2}
    If a proposal with value $v_c$ and round $r_c$ is chosen, then every proposal
    issued with round $r > r_c$ by any proposer has also value $v_c$.
\end{prop}
\begin{proofsketch}
By proposition~\ref{prop:1}, there is a quorum $Q$ that has voted for $v_c$ in $r_c$.
Since any two quorums have a non-empty intersection, any proposer $p$ will receive
at least one \TAGack\ reply of an acceptor included in $Q$. Furthermore, no acceptor has
voted for a proposal valued $v'$ with $v' \neq v$ in round $r'$ with $r' > r_c$.
This would imply the existence of a quorum $Q'$ for which every acceptor has
acknowledged round $r'$ before voting for the proposal in round $r_c$. This contradicts
the existence of $Q$ since acceptors cannot vote for a lower round than they have
previously acknowledged. Therefore, the proposal with the highest round that $p$
receives has value $v_c$. Thus, $p$ issues a proposal with $v_c$.
\end{proofsketch}

Proposition~\ref{prop:2} assumes that rounds can be totally ordered.
However, they are only partially ordered due to our modified negotiation mechanism.
Thus, we must show:
\begin{prop}
For any round number $n$, at most one proposal is issued.
\end{prop}
\begin{proofsketch}
A proposer can only issue a proposal in a round with round number $n$ once it has
received an acknowledgement from a quorum of acceptors with consistent and increased
\Vrack\ with round number $n$. Any acceptor can send at
most one \TAGack\ message in which it has also increased its \Vrack\ to
have round number $n$. Thus, at most one proposer can receive such a quorum to make
a proposal. If incremental rounds are not used, proposers have to choose their own
unique round numbers (cf. canonical Paxos).
\end{proofsketch}
\begin{prop} \label{prop:write-once-linearizable}
The Paxos-based write-once atomic register is linearisable.
\end{prop}
\begin{proofsketch}
Proposition~\ref{prop:1} and~\ref{prop:2} show that all writes return value $v_c$
of the first chosen proposal as their result. Reads differ from writes in
that they can return the initial value $\bot$, but only if no value is chosen since a consistent quorum
is required. Since a proposer must have learned a value before any write (or read
not returning $\bot$) can complete, any subsequent read will return $v_c$.
\end{proofsketch}

\subsection{Consensus Sequence Register} \label{sec:cmd_sequence}
The typical approach to learn a sequence of consensus values is to
chain multiple consensus instances on separate resources~\cite{lamport2001paxos,DBLP:conf/podc/ChandraGR07}.
In contrast, we aim to operate on the same set of resources. For that, we extend
our fault-tolerant write-once atomic register to support a sequence of updates.

The interface of our extended register changes slightly. Instead of including a
specific value \Vval\ (see Algorithm~1, line~1)
in a write request, clients include an update command \Fcmd, which transforms
the current value of the register to the next value. The required changes of
the proposer's second phase are highlighted in Algorithm~2.
The behaviour of the acceptors remains unchanged.

We introduced the concept of consistent quorums in our write-once register
to detect if the current value is chosen or not (see \secref{sec:write-once}).
We can use this information to handle a sequence of updates:
A proposer is allowed to propose a new value if the current value is
chosen. Otherwise, it must complete the unfinished consensus by proposing an
existing value. We refer to the former as a 
\newterm{successor proposal} and to the latter as a \newterm{write-through proposal}.

Line~5--13 shows how a proposer submits
a successor proposal. It first applies the update command \Fcmd\ it has received from the
client on the current established value. If the update is a valid operation,
the proposer can send the result to all acceptors. Sometimes, the update reduces
to a no-op as it cannot be applied to the current value, for instance, if
it includes compare-and-swap semantics or requires a write lock that is missing.
The proposer does not have to complete the second phase as the update has no effect
and can therefore immediately return to the client.

The submission of a write-through proposal (line~14--23)
is equivalent to proposing a value using our write-once register. The proposer
proposes the value seen in the highest round. Afterwards, it must re-execute
the protocol to process the received write request as a successor proposal.

\begin{description}[listparindent=\parindent, topsep=0pt, leftmargin=0cm]
    \item[Safety.]
Intuitively, the register behaves as if executing multiple single-decree Paxos
instances in sequence, with each instance using the previously chosen
proposal and its round as initial state. Updates are applied on top of
a chosen value, which is ensured by observing a consistent quorum. Thus, for
any two values $v_1$ and $v_2$ that are chosen in this order, $s(v_1)$ is the
prefix of $s(v_2)$. By an argument analogous to
proposition~\ref{prop:write-once-linearizable}, reads always return the
latest chosen value. Thus, CS-Stability and CS-Consistency are satisfied.

An update $u$ can only complete if a value that includes $u$ in its update sequence
is chosen, as a quorum of \TAGvoted\ messages is required. As only chosen values
are returned, CS-Update-Visibility is guaranteed.
No proposer applies $u$ on any chosen value after $u$ is
completed. Thus, every subsequent update appears after the last occurrence of $u$
in $s(v)$ of a subsequently chosen value $v$ (CS-Update-Stability).
\end{description}

\begin{figure}[t]
    \vspace{0.36em} 
    \hspace{-0.15cm}
    \includegraphics[width=\linewidth+0.2cm]{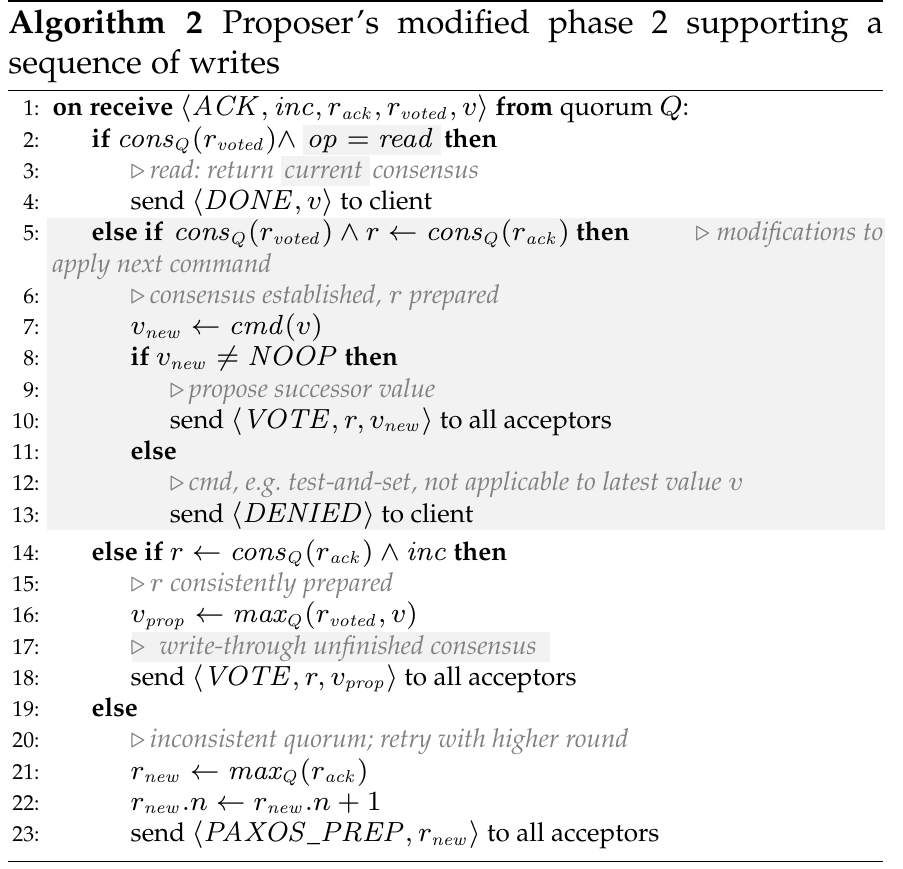}
\end{figure}

\subsection{\approach/: Atomic Read-Modify-Write Register} \label{sec:atomic_cmd_sequence}
\begin{figure*}[t]
    \includegraphics[width=\textwidth]{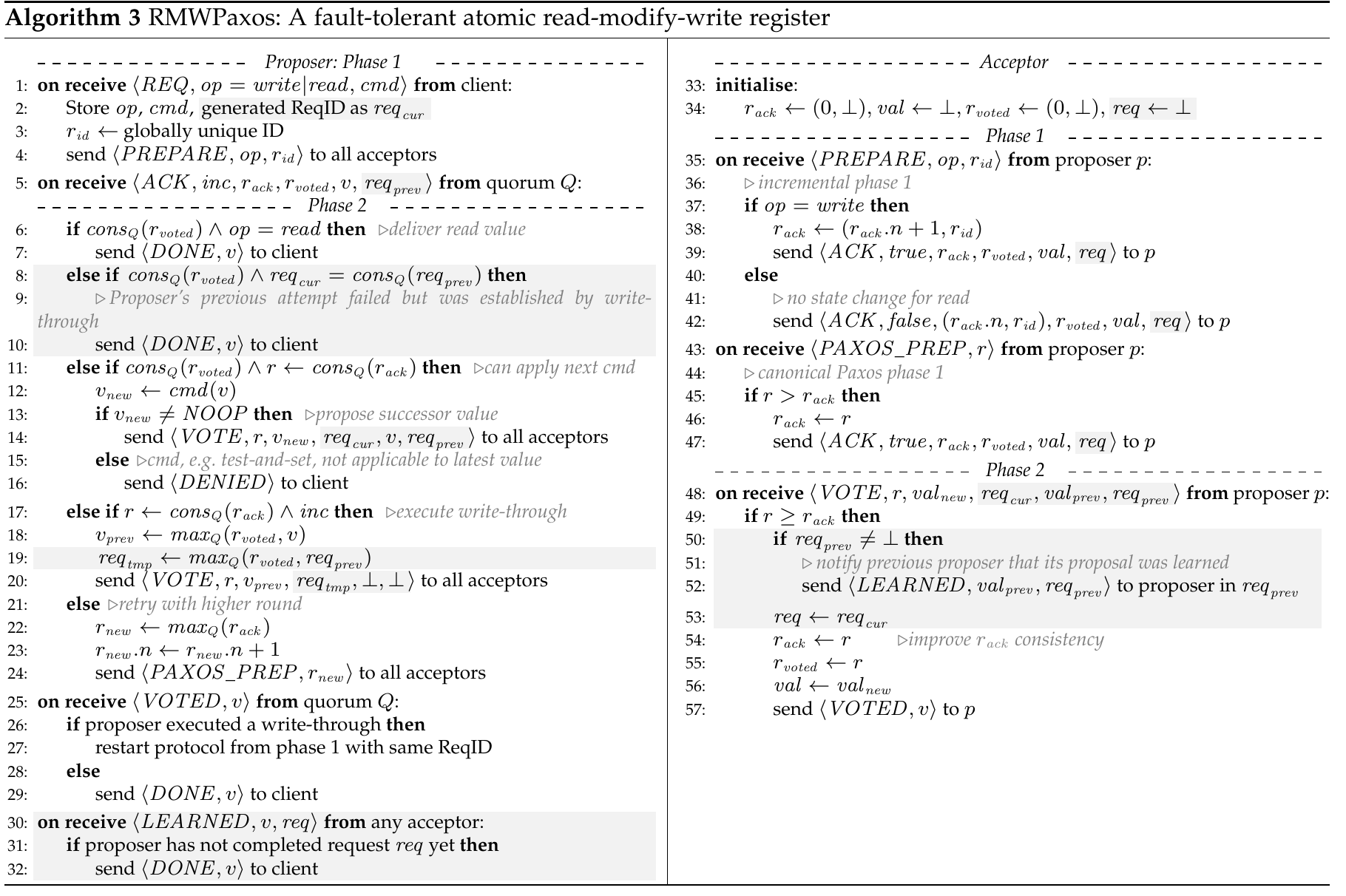}
\end{figure*}

The consensus sequence register presented in the previous section is not atomic,
as it is possible that an update command submitted by a client is proposed and
applied multiple times by the same proposer. For example, consider the following scenario:
Proposer $p_1$ completes phase 1 and submits a successor proposal. However, it only gets a
minority of acceptor votes, as some concurrent proposer $p_2$ already increased the
\Vrack\ rounds of a quorum of acceptors. In this case, $p_2$ may observe an
inconsistent quorum and therefore executes a write-through of $p_1$'s proposal.
If it succeeds, then $p_1$'s proposal was effectively accepted because the value
proposed by $p_1$ is chosen. However, $p_1$ does not know this and retries, potentially
executing the command twice.

For atomicity, we must ensure that a proposer does not re-submit a successor proposal
once the proposed value of a previous attempt is chosen. For that, we assume
reliable in-order message delivery (see \secref{sec:system_model}).
This can be provided by reliable communication protocols such as TCP.
Note, that messages can be lost if a TCP connection fails
and is later re-established during the processing of a request.
To solve this issue, processes can be treated as crashed until the request
is completed. Now, the protocol can be modified as follows (cf. Algorithm~3):

For every write request that proposer $p$ receives, it generates a request ID (ReqID) consisting
of its PID and some locally unique value (line~2). Every acceptor holds the ReqID of the last
proposal it voted for and includes it in all phase~1 \TAGack\ messages it sends.

If a proposer submits a successor proposal, it includes its own ReqID as \Vreq\Pcur\ and the ReqID
received in phase 1 as \Vreq\Pprev\ in its \TAGvote\ messages
(line~14). Here, \Vreq\Pprev\ indicates the last successor proposal that was
chosen by the register. If the proposer submits a write-through, it includes
the ReqID received in phase~1 as \Vreq\Pcur. Since the last chosen proposal is now known,
\Vreq\Pprev\ remains empty (line~20).

Each time an acceptor votes for a new proposal, it updates $\Vreq$ to $\Vreq\Pcur$ (line~53).
If $\Vreq\Pprev$ is non-empty, it sends a \TAGlearned\ message to the respective proposer
(line~52). Receiving a \TAGlearned\ message guarantees that the corresponding proposal
was chosen. A proposer that retries a request with some ReqID can stop the protocol if (1) it observes
a consistent quorum with this ReqID (line~8), or (2) it receives a \TAGlearned\ message with it
(line~30). In both cases, it notifies the client
that its write request succeeded.

We note that it is easy to avoid sending values in \TAGlearned\ and \TAGvoted\ messages back
to the proposer if the proposer keeps track of its proposed values locally. By extension, it is not
necessary to include \Vval\Pprev\ in \TAGvote\ messages.
For simplicity, this is not shown in Algorithm~3.

\begin{description}[listparindent=\parindent, topsep=0pt, leftmargin=0cm]
\item[Safety.]
Assume a write request with ReqID $r$ and update command $u$ is processed by proposer $p$.
Assume that $p$'s attempt failed, but its proposed value is chosen (e.g. due to a write-through).
Proposer $p$ does not propose $u$ as the direct successor of its own proposed value because
it would observe a consistent quorum with ReqID $r$ beforehand. Thus, assume
that some successor value proposed by a different proposer is chosen.
This means that \TAGlearned\ messages with ReqID $r$ are sent to $p$ by some quorum $Q$.
Let $p$ retry its request. In order to apply $u$ and
propose a new value, $p$ must observe a consistent quorum $Q'$.
As $Q \cap Q' \neq \emptyset$ and reliable ordered links are used, $p$ receives
a \TAGlearned\ message before receiving a consistent quorum.
Thus, $p$ does not apply $u$ on a value whose update sequence already includes $u$.
\end{description}

\subsection{State Machine Replication}\label{sec:RSM}
By using \approach/, we can build a fault-tolerant
replicated state machine using a fixed set of storage resources.
The state is stored in the register and state changes are done by the
corresponding update commands. If updates are idempotent, the
consensus sequence register suffices. One way to achieve this is by
using transactional semantics such as compare-and-swap.

In log-based approaches like Multi-Paxos~\cite[Sect.~3]{lamport2001paxos},
acceptors accept commands, i.e. state transitions of the state
machine. In our approach, in contrast, the acceptors accept the complete state. This has several implications. First, a dedicated set of learner processes is no longer required.
Any process that wishes to learn the current state of the RSM can do so by
executing a read. This process then acts as the sole learner in the context of
this command. In contrast, Multi-Paxos requires multiple learners in order to
have access to the state in a fault-tolerant manner. Since every learner must also
learn every command to make progress, $n*m$ \TAGvoted\ messages are required
in a setup with $n$ acceptors and $m$ learners.
Our approach requires only $n$ \TAGvoted\ messages.

Second, by keeping the full state in acceptors, a sequence of commands can now be applied
to the RSM in-place using the same set of acceptors. Thus, it is not necessary to
allocate and free storage resources. This simplifies the
protocol's complexity and its implementation. Due to the absence of any state management
overhead, it is trivial to use arbitrary many \approach/ instances in parallel, allowing
a more fine-granular use of the RSM paradigm. This is especially useful if the
state can be split into many independent partitions, as
it is often the case in key-value structured data.

\subsection{Liveness}
Reads and writes are obstruction-free~\cite{DBLP:conf/icdcs/HerlihyLM03} as long
as a quorum of acceptors and the proposer receiving the requests are correct.
Wait- or lock-freedom~\cite{DBLP:journals/toplas/Herlihy91} cannot be guaranteed without
further assumptions, as postulated by the FLP result~\cite{DBLP:journals/jacm/FischerLP85}.
A common assumption is the existence of a stable leader to which all
requests are forwarded. The leader then acts as the sole proposer of the system.
To handle leader failures, a $\Diamond\mathcal{W}$ failure detector~\cite{DBLP:journals/jacm/ChandraHT96}
is necessary.

\subsection{Optimisations} \label{sec:optiseqwrite}
There are several ways to optimise the basic protocol.

\begin{description}[listparindent=\parindent, topsep=0pt, leftmargin=0cm]
\item[Fast Writes.] Handling writes requires a proposer to complete both phases
    of the protocol. That means that at least four message delays are needed. By
    using a mechanism similar to Multi-Paxos~\cite{lamport2001paxos}, the first phase
    can be skipped by a proposer that processes multiple writes uninterrupted
    by other proposers.  We refer to such writes as \newterm{fast writes}.

    The modification is simple. Whenever an acceptor votes for a proposal made in
    round $r$, it sets \Vrack\ to $(r.n + 1, r.\mathit{id})$
    (cf. Algorithm~1 line~40). By doing
    so, it effectively behaves as if receiving a \TAGprep\ message from
    the same proposer immediately after voting. Therefore, this proposer can skip
    the first phase when making its next proposal.

    This optimisation is useful for single-writer settings or scenarios in
    which a proposer must execute multiple writes within a short period.
    As no locking or lease mechanism is used, an ongoing fast write sequence can
    be interrupted at any time by other proposers. Thus, we avoid the costs
    and unavailability associated with a leader and its (re-)election.

\item[Fewer concurrency conflicts caused by reads.]
If a read observes a consistent quorum after the first phase, it returns
a result without interfering with any concurrent request because acceptors
do not modify their rounds. If a read observes an inconsistent quorum, a write-through
is triggered, which can cause interference. Write-throughs cannot be prevented
completely, as a crashing proposer can cause a proposal to be only partially established.
Therefore, we adopt the idea of contention management~\cite{DBLP:conf/icdcs/HerlihyLM03,DBLP:conf/podc/SchererS05}
to unreliably detect a crashed writer:

When a reading proposer observes an inconsistent quorum, it stores the highest
round it has received. Then, it retries phase 1 without an explicit round. If the quorum is
again inconsistent, it checks whether progress was made by comparing the received
rounds with the rounds from the previous iteration. If they remain unchanged,
then it is possible that the write crashed and a write-through must be triggered.
Otherwise, the reader can try again. The proposer can keep collecting replies
from its previous attempts as it is possible to reach a consistent quorum with delayed
replies.

To prevent a read from starving due to a continuous stream of writes, we define
an upper limit on the number of retry attempts. Its effects are
evaluated in \secref{sec:practical-evaluation}.

\item[Batching.] Batching is a commonly used engineering technique to reduce bandwidth
    and contention by bundling multiple commands in a single request at the cost
    of higher response latency. Every proposer manages separate batches for read and update commands.
    A batch is processed at regular intervals by starting the protocol. For write
    batches, all update commands of the batch are applied in-order on the old value before
    proposing the resulting new value. When processing a read batch, the read value is simply
    returned to all clients. The size of all messages remains constant, independent
    of the number of batched commands. This shifts the performance bottleneck from
    internal communication to the processing speed of the respective proposers.

\end{description}

\begin{figure*}[t!]
    \centering
    \captionsetup{justification=centering}
    \begin{subfigure}[t]{\textwidth}
        \centering
        \includegraphics[draft=false,width=\textwidth]{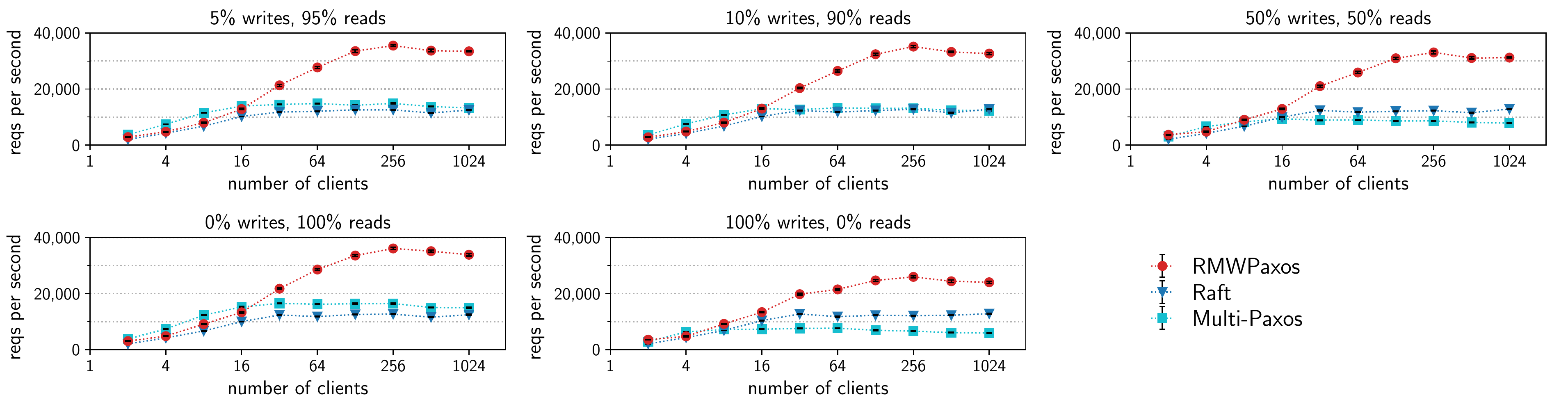}
        \caption{Throughput comparison with an increasing number of clients}
        \label{fig:tp_compare_load}
    \end{subfigure}%
    \\ \vspace{1em}
    \begin{subfigure}[t]{\textwidth}
        \centering
        \includegraphics[draft=false,width=\textwidth]{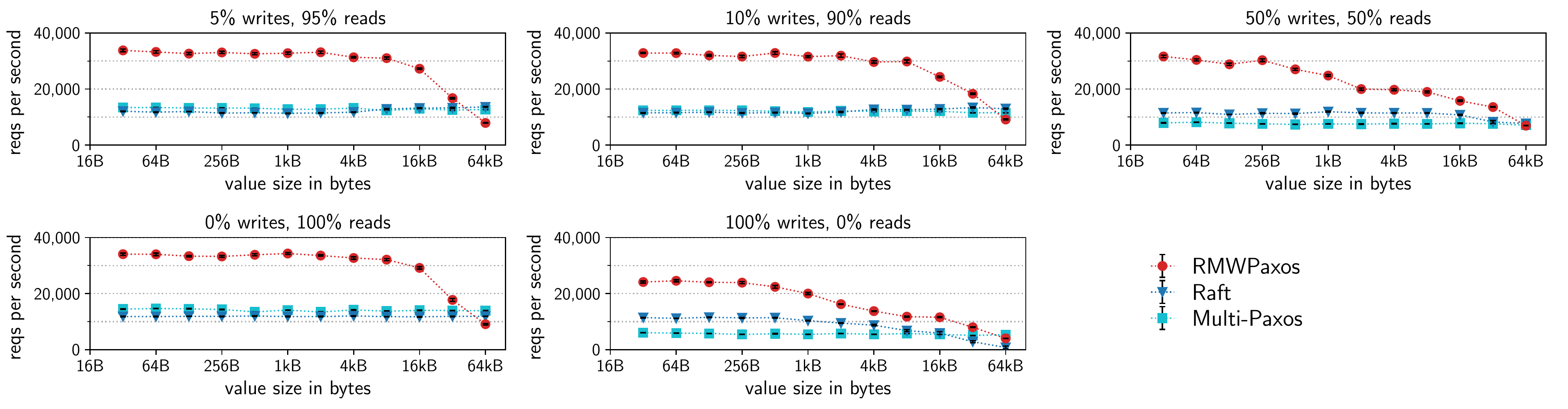}
        \caption{Throughput comparison with increasing value sizes (using 512 clients)}
        \label{fig:tp_compare_size}
    \end{subfigure}%
    \vspace{2mm} 
    \caption{Comparing the throughput of \approach/ with Raft and Multi-Paxos
    using three replicas.}
\end{figure*}

\section{Analysis}\label{sec:evaluation}
In this section, we focus on additional aspects that might be beneficial for practical
deployments. An experimental evaluation can be found in \secref{sec:practical-evaluation}.

Compared to canonical Paxos and Multi-Paxos our registers require a
similar number of 2--4 message delays per consensus in the conflict-free case.
Two additional message delays are needed by canonical Paxos when a
valid round number is not known yet and by our registers when a read
using incremental rounds has to help to establish a
consensus. Reading a stable, established consensus with our approach
only needs 2 message delays, no concurrency control and does not cause
acceptor state changes, which is costly if their state must be persisted.
Furthermore, our approach
works on a fixed set of resources which makes dynamic resource
allocation, pruning, and deallocation needless.  This makes our
register applicable on a more fine-granular level than other
consensus-based approaches that rely on a command log.

Relying on consistent quorums does not harm robustness nor
performance. Like in canonical Paxos, a single replica with the
highest round seen in an inconsistent quorum will suffice to propose
its value.  But on a \emph{consistent quorum}, we can (a) terminate a
read operation early by not needing to write and re-learn the
consensus and (b) can base the next consensus in our consensus
sequence on that.

Not requiring an explicit leader provides more continuous availability.  In
our approach, any proposer can issue requests to the register at any
time. When a proposer fails, other proposers can immediately proceed
and do not need to wait for or elect a new leader. Still, a proposer
submitting many requests in sequence without any interference of other
proposers can perform each write to the register in just two message
delays (no batching), like a leader.

\section{Experimental Evaluation}

\label{sec:practical-evaluation}

We implemented \approach/ in Scalaris~\cite{DBLP:conf/erlang/SchuttSR08}, a distributed
key-value store written in Erlang. The correctness of our implementation was tested using a protocol
scheduler~\cite{scalaris_proto_sched}, which forces random interleavings of incoming messages.
We detected no safety violations using this approach.

The primary focus of the evaluation is to show the scalability of
our approach under different workloads, as absolute performance is highly
dependent on the available hardware environment and engineering efforts that are
independent of the actual approach. Our register aims to be a general
primitive. Thus, we consider use-case dependent techniques
that optimise network traffic and concurrent access, e.g. request batching,
being out-of-scope of this paper.

All benchmarks were performed on a cluster with two Intel Xeon E5-2670
v3, 2.40\,GHz per node. All nodes are fully-connected with 10\,Gbit/s
links. Each cluster node hosts a single replica,
which is a Scalaris node that encapsulates one proposer and one acceptor process. Load generation
was performed on up to two separate cluster nodes using the benchmarking tool Basho
Bench~\cite{github_basho_bench}, which was modified to enable workloads with heterogeneous
client processes.  In all experiments, Basho Bench clients were distributed evenly
across the load generating nodes. All clients submit their requests sequentially,
i.e. each client waits for a response before issuing the next request.

All shown measurements ran for 10 minutes with request data aggregation
in one-second intervals. We show the mean with 99\,\% confidence intervals (CI)
and 99th percentile latencies. In almost all cases, the CI lies within two percent of the reported median.

\begin{figure*}[t!]
    \centering
    \captionsetup{justification=centering}
    \begin{subfigure}[t]{0.32\textwidth}
        \centering
        \includegraphics[draft=false,width=\textwidth]{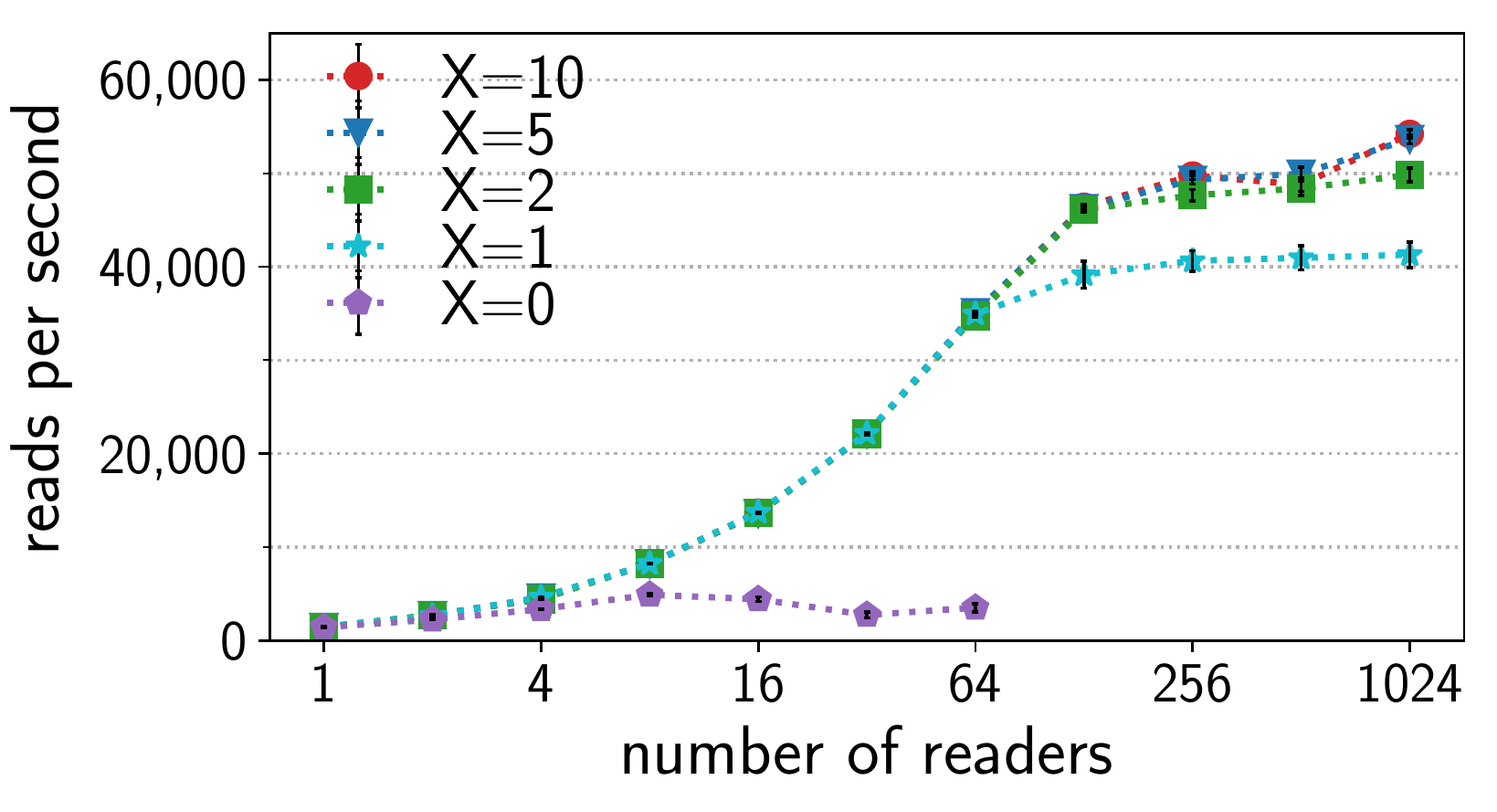}
        \caption{Read throughput}
        \label{fig:single_tp_read}
    \end{subfigure}%
    ~
    \begin{subfigure}[t]{0.32\textwidth}
        \centering
        \includegraphics[draft=false,width=\textwidth]{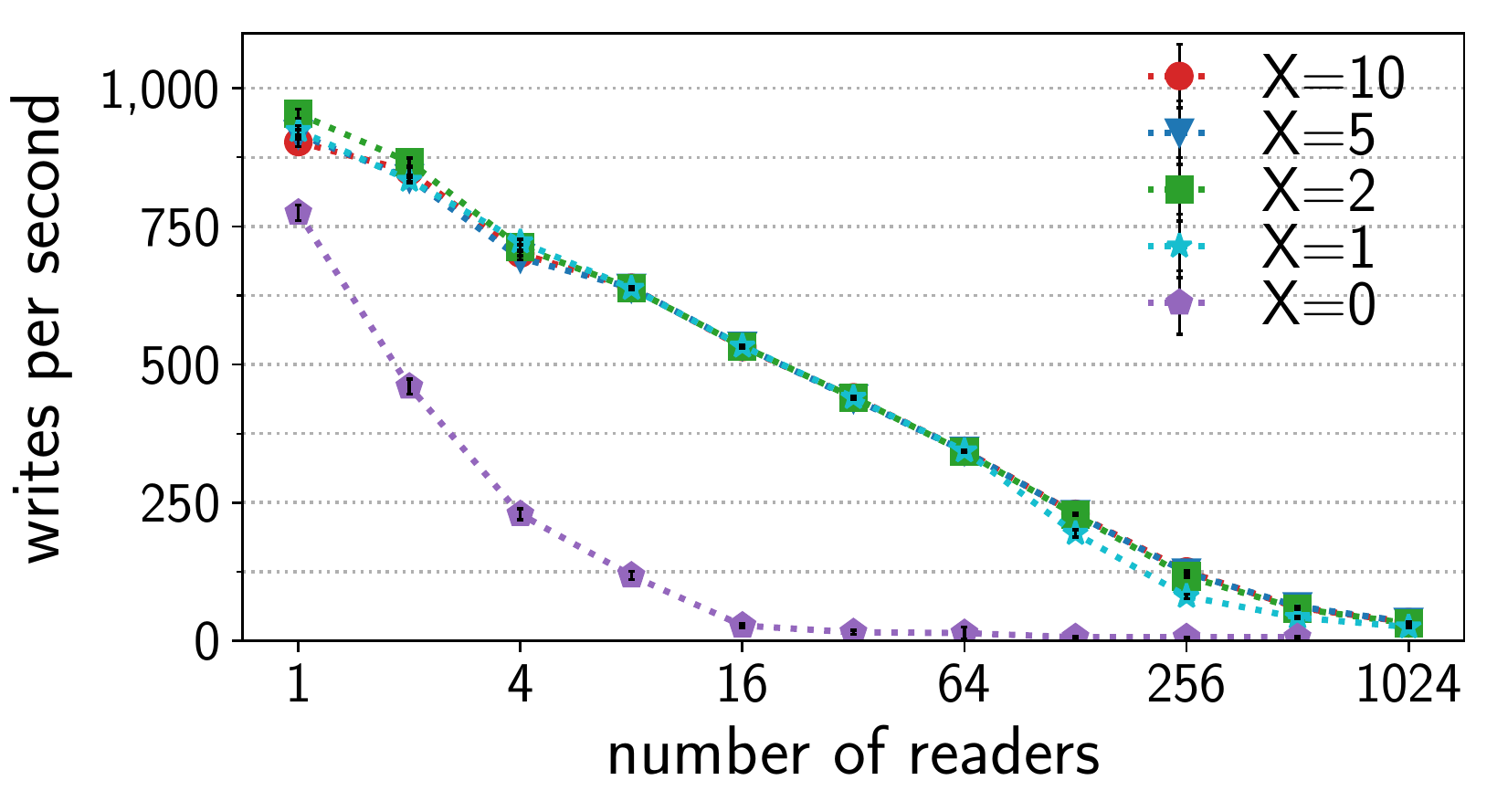}
        \caption{Write throughput}
        \label{fig:single_tp_write}
    \end{subfigure}
    ~
    \begin{subfigure}[t]{0.32\textwidth}
        \centering
        \includegraphics[draft=false,width=\textwidth]{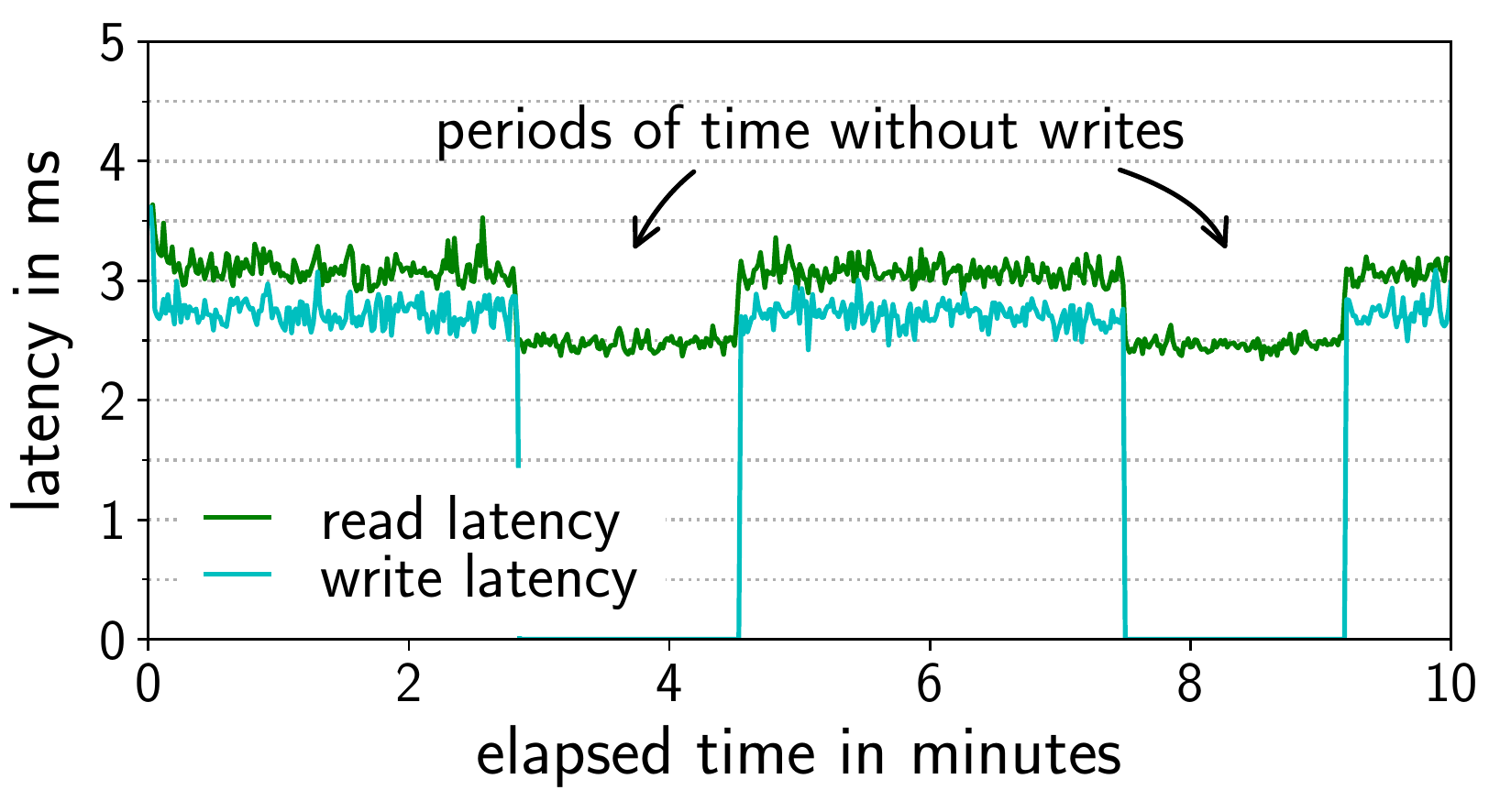}
        \caption{99th pctl. latency (64 readers, \textsf{X}=10)}
        \label{fig:single_lat_hanging}
    \end{subfigure}
    \vspace{1mm} 
    \caption{Single-writer performance of \approach/ with five replicas.}
    \label{fig:single_writer_multiple_reader}
\end{figure*}
\begin{figure*}[t!]
   \vspace{-3mm}
    \centering
    \captionsetup{justification=centering}
    \begin{subfigure}[t]{0.319\textwidth} 
        \centering
        \includegraphics[draft=false,width=\textwidth]{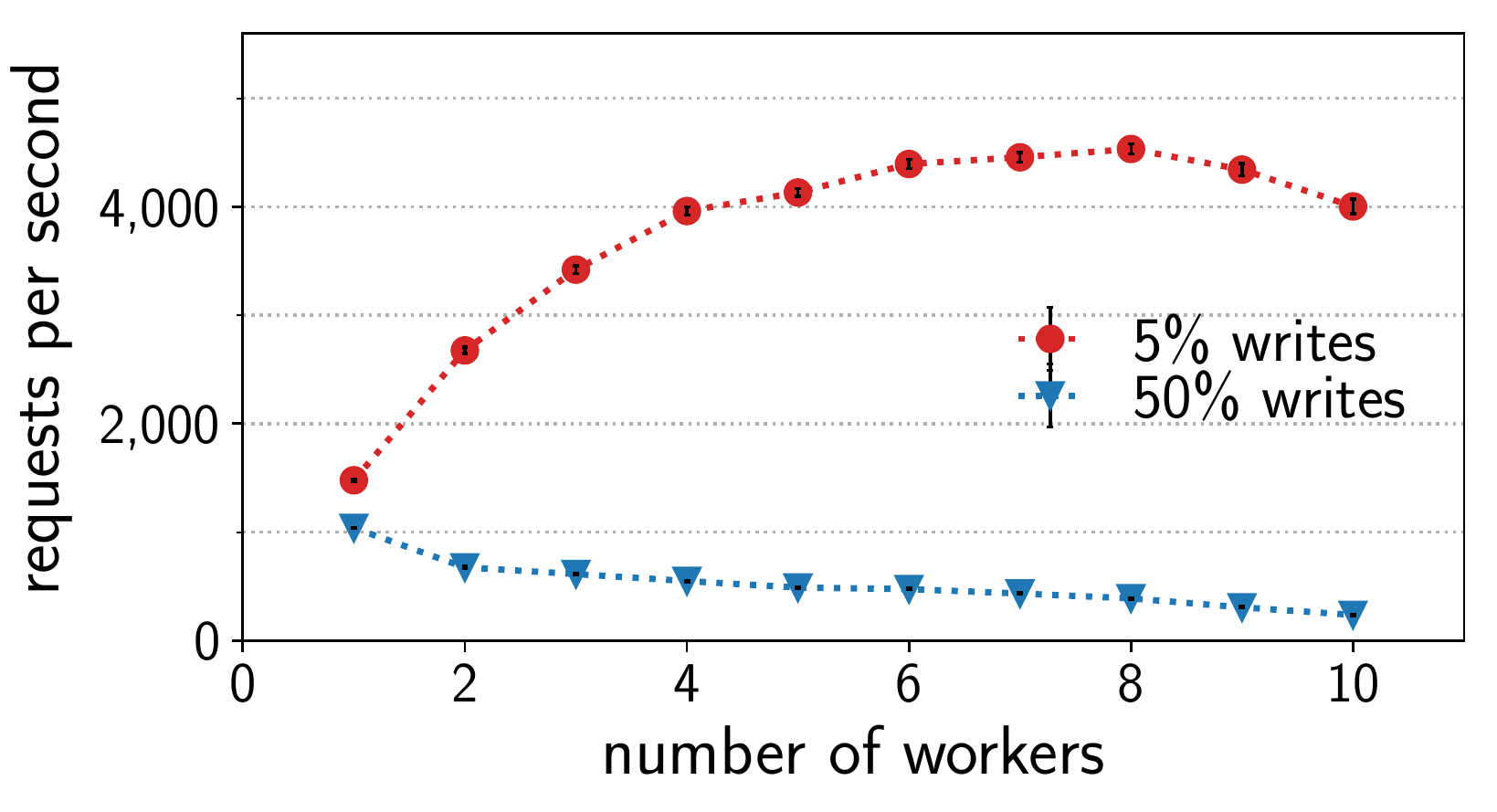}
        \caption{Single register throughput}
        \label{fig:multi_single_reg}
    \end{subfigure}%
    ~
    \begin{subfigure}[t]{0.32\textwidth}
        \centering
        \includegraphics[draft=false,width=\textwidth]{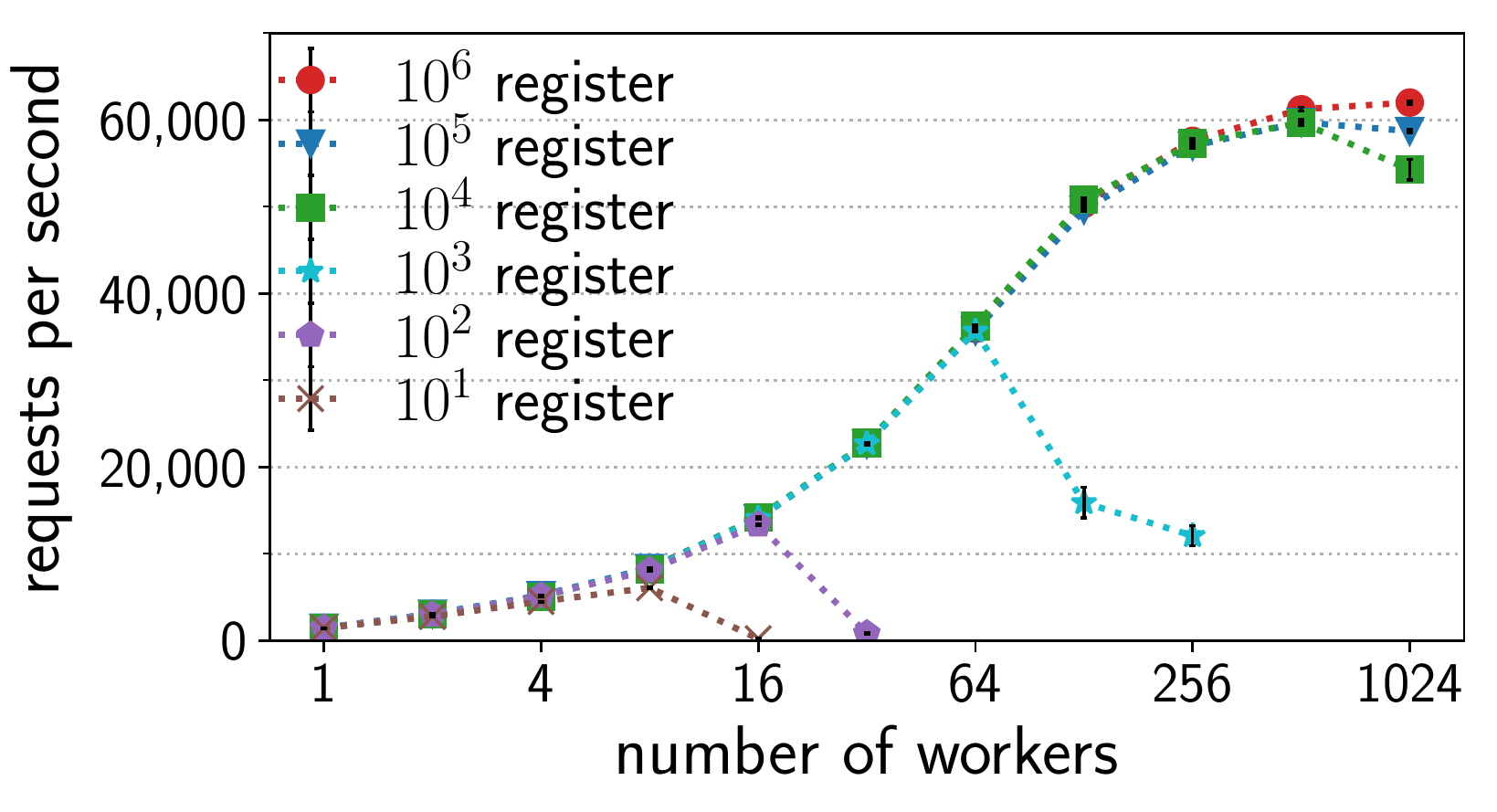}
        \caption{Multi-register read-heavy throughput}
        \label{fig:multi_reg_read}
    \end{subfigure}
    ~
    \begin{subfigure}[t]{0.322\textwidth} 
        \centering
        \includegraphics[draft=false,width=\textwidth]{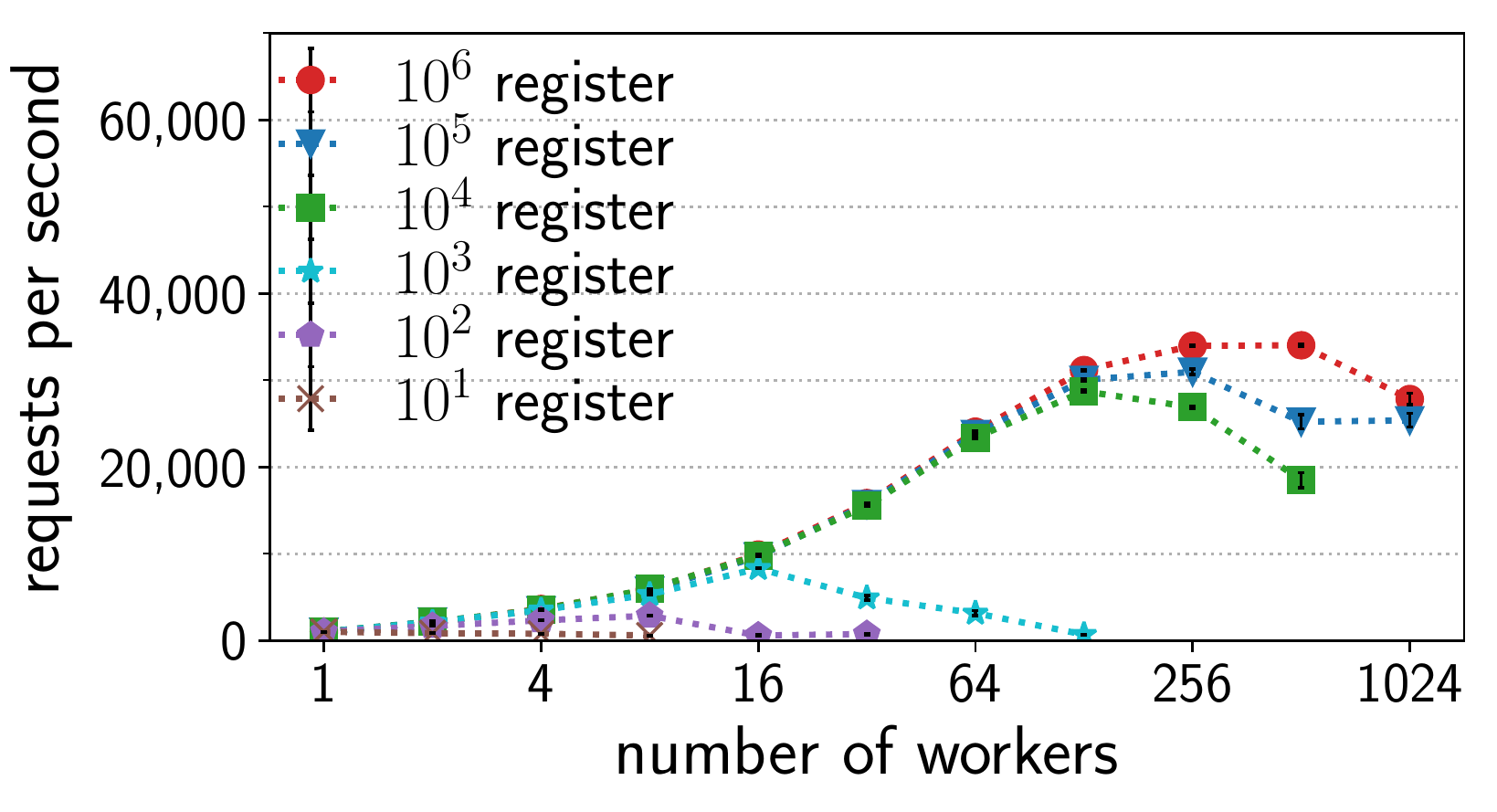}
        \caption{Multi-register write-heavy throughput}
        \label{fig:multi_reg_write}
    \end{subfigure}
    \vspace{1mm} 
    \caption{Leaderless multi-writer performance of \approach/ with five replicas.}
    \label{fig:multiple_writer_multiple_reader}
   \vspace{-1mm}
\end{figure*}

\subsection{Comparison with Raft and Multi-Paxos}
First, we compare the performance of \approach/ with open-source implementations of
Multi-Paxos~\cite{github_riak_ensemble, lamport2001paxos} and
Raft~\cite{github_raft, DBLP:conf/usenix/OngaroO14}, two commonly used state-of-the-art protocols.
To minimise the performance impact of the IO subsystem, we configured both approaches
to write their data to RAM disk. In \approach/, data is stored by using
Erlang's build-in term storage~\cite{erlang_ets}. All approaches use three replicas.
As both Multi-Paxos and Raft make use of a leader, we simulate a leader by randomly selecting
one node to which all requests are forwarded to in the case of \approach/. As any leader election
protocol can be implemented on top of \approach/, we consider leader election to be
out-of-scope.

We measured the throughput of all approaches in scenarios: First, a counter
that is accessed by an increasing number of clients (\figref{fig:tp_compare_load}).
Second, a binary value of increasing size accessed by a fixed number of clients (\figref{fig:tp_compare_size}).

All three approaches handle requests in a single round-trip between leader and
a quorum of following nodes. Thus, the observed differences can largely be attributed
to their different strategies in handling the data locally. Due to the absence of
any state management, \approach/ consistently outperforms both the Raft and Multi-Paxos implementation
for small state sizes. For the latter two, overhead caused by reading/writing data to the local file system
increases request latency, which in turn negatively affects throughput. In addition, the Multi-Paxos and
Raft implementations use mechanisms such as checksum validation to protect against disk corruption.

We note that Multi-Paxos has a higher throughput than \approach/ in read-heavy
workloads with few clients. We attribute this to our method of load generation. As clients submit requests
sequentially, both approaches do not reach full capacity. Here, we observe a slightly
lower mean read latency for Multi-Paxos (0.6ms vs 0.8ms), which is likely caused by
implementation-specific overhead.

For values smaller or equals to 4kB, all approaches exhibit nearly constant read performance. However,
the throughput of \approach/ decreases for larger values. This is because the full value is always transferred
from a quorum of nodes to the proposer when executing a read. This causes high communication costs
in settings where individual objects have moderate or large size. However, analysis of existing large-scale key-value
stores have shown a heavy skew towards small values of less than a kilobyte~\cite{DBLP:conf/sigmetrics/AtikogluXFJP12, DBLP:conf/nsdi/NishtalaFGKLLMPPSSTV13}.

In contrast to \approach/, the Raft and Multi-Paxos implementations include optimisations to keep data
transfer costs between nodes constant when executing a read if the leader is stable.
In Raft's case, an empty heartbeat log entry must be appended to the command logs to
ensure that the data of the leader is up-to-date.
This introduces a slight overhead when reading entries.

\subsection{Leaderless Performance} \label{sec:leaderless}
\approach/ is derived from Paxos. Thus, it does not depend on the existence of a leader
to satisfy the safety properties of consensus, in contrast to protocols like Raft, which
do not work without a single leader. However, a leader
is beneficial for progress because it prevents the duelling proposer problem.
For \approach/, we can alleviate the need for a leader as it is trivial
to deploy an arbitrary number of concurrent \approach/ instances. This way,
load on a single instance can be greatly reduced, depending on the workload.

We examined both single-writer (\figref{fig:single_writer_multiple_reader})
and multi-writer (\figref{fig:multiple_writer_multiple_reader}) workloads, as
previous work in the design of data structures has shown that supporting
concurrent modifications often inhibits their performance~\cite{DBLP:conf/podc/SchererS05}.
To better illustrate the effects of concurrent requests, we increased the system
size to five replicas (acceptors).

\begin{description}[topsep=0pt, leftmargin=0cm, listparindent=\parindent]
\item[Single-Writer.] To evaluate single-writer performance, we used
one writing client and up to 1024 concurrent readers
with a different number of read retries (parameter \textsf{X}). The results are depicted
in \figref{fig:single_writer_multiple_reader}.
We observed that even a single retry (\textsf{X}=1) improves both read and write
throughput greatly compared to disabling this optimization (\textsf{X}=0). In the latter
case, the register was overloaded due to concurrent write-through attempts by the readers
if more than 64 readers where used, dropping throughput to 0 at some times.
As these results are not stable, they are not shown in \figref{fig:single_tp_read}. Choosing
a value for \textsf{X} larger than 2 has only a minor impact on the read throughput.
As acceptors must handle more messages with an increasing number of clients, their response
latency increases. This leads to a consistent decline of the write throughput, as shown
in \figref{fig:single_tp_write}.
Since the load is distributed more evenly across all replicas, the maximum observed
throughput increased by roughly $70\,\%$ compared to our leader-based experiments (cf.
\figref{fig:tp_compare_load}), even though the system size increased from 3 to 5 replicas.

\figref{fig:single_lat_hanging} shows the latency impact of using
read retries. Read latency only increases by approx. 0.5\,ms in the presence of a concurrent
writer. This may contradict the expectation that some reads require multiple 
round trips as they can observe an inconsistent quorum initially. However, proposers
can continue collecting replies from the initial attempt and return a result once they
observe a consistent quorum. As there is only a single writer, such a quorum always exists,
at the latest after receiving a reply from every acceptor.  This also
means that reads trigger no write-throughs. Thus, both reads and writes succeed
after a single round trip in a single writer setup as long as no acceptor fails. Note that
writes exhibit a slightly lower latency as they always succeed with a quorum of replies,
wheres reads must potentially wait for all replies in some cases.

\item[Multi-Writer, Single-Register.]
All clients sent a uniform mix of read and write requests for
the evaluation of multiple writers. \figref{fig:multi_single_reg} compares
the throughput of a read-heavy workload (5\,\% writes) with a write-heavy
workload (50\,\% writes)~\cite{DBLP:conf/cloud/CooperSTRS10}. Performance degradation caused by duelling proposers
can be observed for both workloads. The throughput of the read-heavy workload scales
up until four concurrent clients. Afterwards, clients begin to invalidate each
other's proposals repeatedly. In write-heavy workloads,
even two concurrent clients are enough to have a negative impact on the system's
performance. As shown in the previous experiments, a leader at the application
level helps to handle write concurrency effectively.

\item[Multi-Writer, Multi-Register.]
All previous measurements focused on a single register. As highlighted in \secref{sec:RSM},
the absence of state management overhead easily allows for arbitrary many registers to be
used. We benchmarked configurations using up to $10^6$ register instances and 512 concurrent
clients. The registers were accessed according to a Pareto distribution~\cite{Newman_2005} with
$\alpha\approx1.16$ ($80\,\%$ of requests targeted $20\,\%$ of registers).
Figures \ref{fig:multi_reg_read} and \ref{fig:multi_reg_write} show the results
for read-heavy (5\,\% writes) and write-heavy (50\,\% writes), respectively. The results
are as expected. More concurrent clients can be handled without performance degradation
due to duelling proposers by increasing the number of parallel registers. The load is evenly
distributed across all replicas, as no leader is used. In addition, contention is low
in settings with a large number of parallel registers. This results in a higher achievable
throughput than it is possible with the use of a leader (cf. \figref{fig:tp_compare_load}).

\approach/ performs consistently better under read-heavy workloads, which coincides
with the results from the single-register evaluation. We used
the read-write ratios of YCSB~\cite{DBLP:conf/cloud/CooperSTRS10}, a benchmarking framework
that aims to simulate real-world use-cases. Studies of large-scale distributed systems
have shown an even higher skew towards reads, reporting read-write ratios of
up to 450:1~\cite{DBLP:conf/sigmetrics/AtikogluXFJP12,DBLP:conf/osdi/CorbettDEFFFGGHHHKKLLMMNQRRSSTWW12, wikimedia_stats}.
\end{description}

\subsection{Impact of Replication Degree and Failures}
\begin{figure}[t]
  \begin{subfigure}[t]{\columnwidth}
    \centering
    \includegraphics[draft=false,width=\columnwidth]{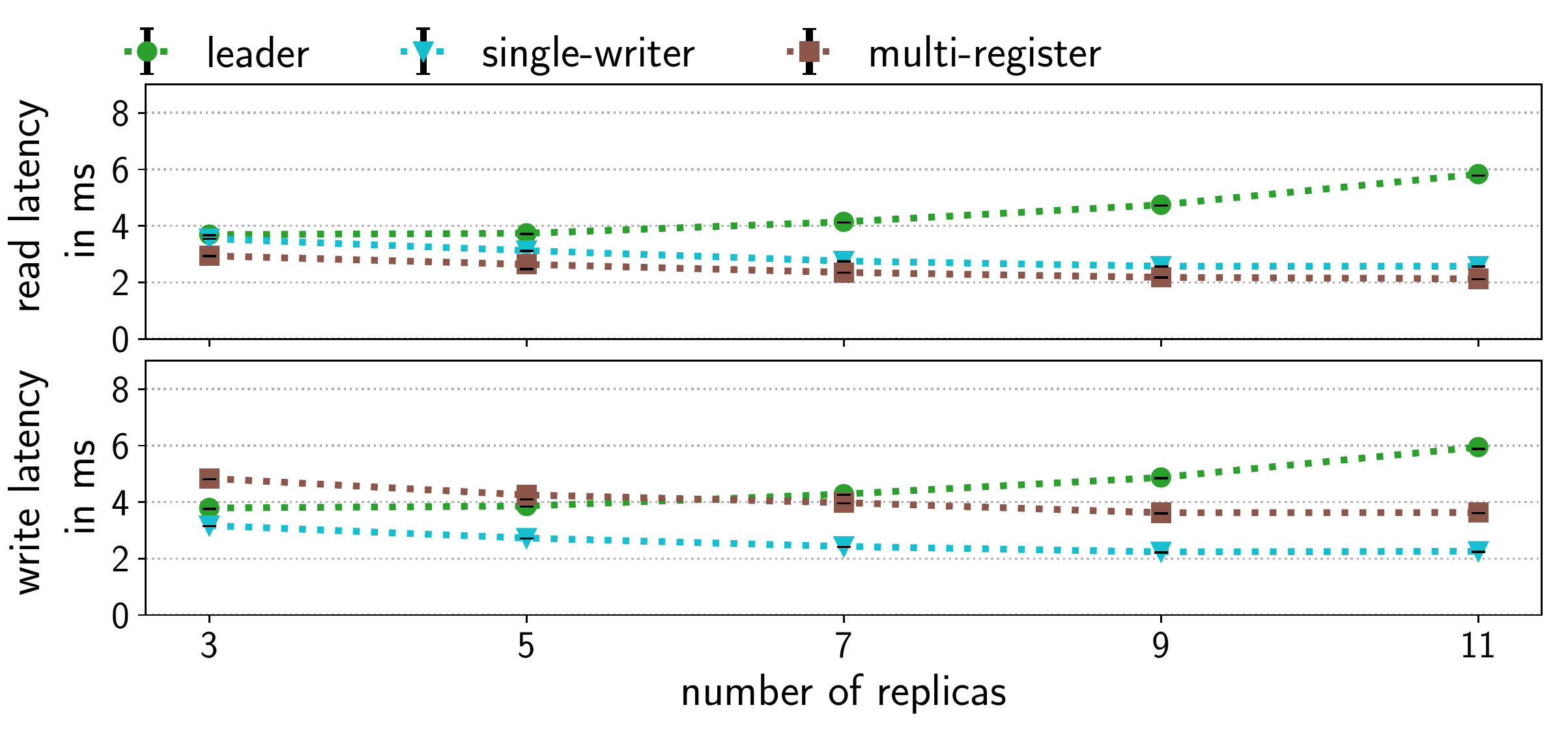}
    \caption{Latency with a growing number of replicas.}
    \label{latsize}
  \end{subfigure}%
  \\
\begin{subfigure}[t]{\columnwidth}
  \centering
  \includegraphics[draft=false,width=\columnwidth]{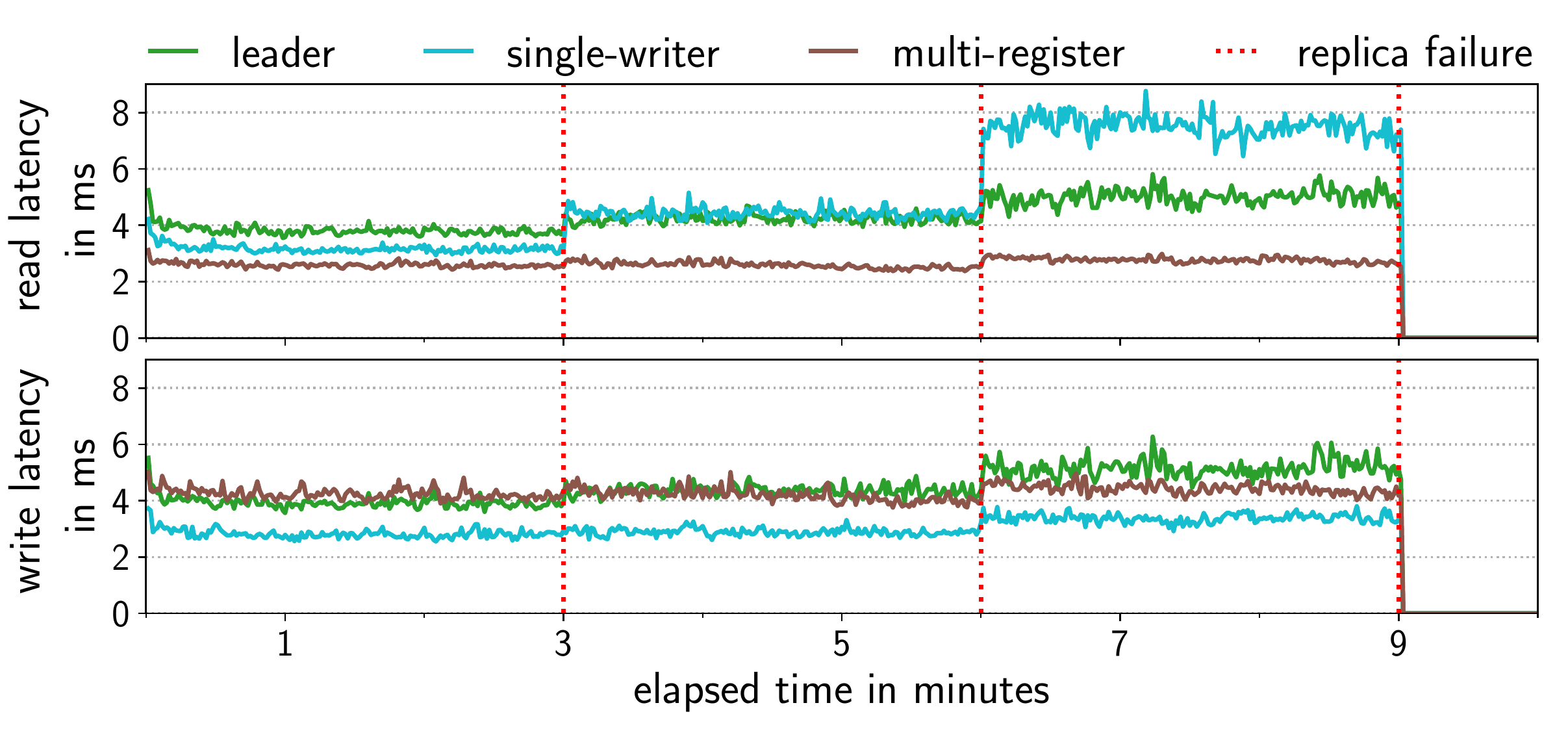}
  \caption{Latency with five replicas and process failures.}
  \label{latcrash}
  \end{subfigure}%
  \caption{99th percentile latency comparison using 64 clients.}
  \label{latcompare}
\end{figure}

We investigated the impact of the number of replicas on the response latency
of \approach/, as well as its ability to tolerate replica failures.
For that, we used different deployment strategies from our previous experiments:
(1) A single register accessed by a leader, (2) a single register accessed by
a single writer and multiple readers, and (3) a 10.000 register setup accessed
by multiple writers and readers with the Pareto distribution used in Section 7.2. 
We will refer to them as the leader, single-writer, and multi-register strategy,
respectively. All measurements were executed using 64 clients. Clients used a
read-heavy workload (5\% writes, 95\% reads) in the leader and multi-register deployment.
The results are shown in \figref{latcompare}.

When using a leader, the number of messages the leader must process increases
with a growing number of replicas. This results in an increasing response latency
as shown in \figref{latsize}. In contrast, the load is distributed evenly among all
node in both the single-writer and multi-register setup. Assuming a constant
throughput, the number of messages each proposer is sending is independent of the
system size. As only replies from a quorum of replicas is needed, fewer messages
must be received in total by each proposer to answer all requests. This results
in a slightly lower response latency of these strategies with growing system sizes.

To measure the impact of failures, we let one replica crash after every three minutes.
Overall, all latencies with the exception of the read latency of the single-writer
strategy remained fairly consistent as long as a sufficient number of replicas is
available. We observed only a slight increase
for each new failure, as proposers must potentially wait for the replies of slower
acceptor processes. However, to ensure that reads can be processed in the
single-writer setup, answers from all replicas are necessary
(see single-writer evaluation in \secref{sec:leaderless}). If a replica fails,
proposers do not always observe a consistent quorum after all remaining acceptors
reply. They must therefore retry their request. This is more likely to happen as
more replicas fail.

\subsection{Leader Load and Applicability to NVM}
Our results show potential for future improvements. First, we aim to improve
the issue of high write contention on a single register while alleviating the
bottleneck caused by a leader. As a single register is able to handle high read
concurrency (see \secref{sec:leaderless}), only writes have to be forwarded to the leader.
This can be coupled with a dynamic leader allocation for only highly contentious registers,
which further reduces the load placed on the leader.

Second, we believe that the fine-granular nature of our approach is a promising fit
for the use in combination with byte-addressable, non-volatile main memory.
With the recent availability of NVRAM, along with the current work
in NVMe over Fabrics~\cite{minturn2015under}, we believe that our approach can
leverage these technologies in the future.

\section{Related Work}
Starting with Lamport's work on the discovery of the Paxos
algorithm~\cite{lamport2001paxos,DBLP:journals/tocs/Lamport98},
numerous Paxos extensions~\cite{DBLP:conf/sosp/MoraruAK13,DBLP:conf/cloud/WangJCYC17,DBLP:journals/dc/GafniL03,DBLP:conf/dsn/MarandiPSP10}
have been proposed---most of them following the design of using multiple Paxos instances
to learn a sequence of commands. As a notable exception, Generalized Paxos~\cite{lamport2005generalized}
and its derivatives~\cite{DBLP:conf/srds/SutraS11}, only use a
single Paxos instance but require keeping track of an ever-growing set of commands
in its messages. In all cases, pruning in some form must be implemented to prevent
unbounded memory consumption, which introduces a considerable amount of complexity to
the system. This is identified by Chandra et al.~\cite{DBLP:conf/podc/ChandraGR07} as one of the main challenges
for using Paxos-based designs in practical systems. Despite numerous efforts of making
Paxos more approachable~\cite{kirsch2008paxos,DBLP:journals/sigact/BoichatDFG03,DBLP:journals/csur/RenesseA15},
reliable state management with Paxos is seldom discussed in detail.
Only a few practical Paxos-based systems exist to this date such as
Chubby~\cite{DBLP:conf/osdi/Burrows06}, Spanner~\cite{DBLP:conf/osdi/CorbettDEFFFGGHHHKKLLMMNQRRSSTWW12},
Megastore~\cite{DBLP:conf/cidr/BakerBCFKLLLLY11}, and Scalaris~\cite{DBLP:conf/erlang/SchuttSR08}.

In recent years, various proposals were made to alleviate the dependence on a single leader.
Mencius~\cite{DBLP:conf/osdi/MaoJM08} evenly shares the leader's responsibilities by
assigning individual consensus instances
to single replicas. In Egalitarian Paxos~\cite{DBLP:conf/sosp/MoraruAK13}, the replica receiving
a command is regarded as its command leader. Each replica can act as a
leader simultaneously for a subset of
commands. This is achieved by decoupling command commit and application from
each other and making use of the dependency constraints of each
command. In contrast, we do not need an explicit leader depending on workload
and load-distribution.

As of today, few consensus protocols, which are not Paxos-based, exist. Most prominently
Raft~\cite{DBLP:conf/usenix/OngaroO14} and the closely related Zab protocol~\cite{DBLP:conf/dsn/JunqueiraRS11}.
Both are based on the idea of a central command log. Furthermore, they
require a \emph{strong} leader, meaning that at most a single leader is allowed to
exist at any given time. In contrast, we perform updates on a distributed state in-place and
do not need a strong leader.

To the best of our knowledge, we present the first Paxos-based approach that does not rely on
additional state management without requiring a leader to satisfy the safety properties
of consensus by implementing an atomic RMW register.
The register by Li et al.~\cite{DBLP:conf/srds/LiCAA07} only recasts the original Paxos
without modification and provides a regular
write-once register. The round-based register proposed by Boichat et al.~\cite{DBLP:journals/sigact/BoichatDFG03} is
not atomic and only write-once. It is similar to the approach of Li et al. and modular
to build several, known Paxos variants such as Multi-Paxos or Fast Paxos~\cite{DBLP:journals/dc/Lamport06}.
CASPaxos~\cite{DBLP:journals/corr/abs-1802-07000} provides a Paxos-based linearisable
multi-reader multi-writer register by letting clients submit a user-defined function
instead of a value. However, when handling concurrent writes it is not guaranteed that
all (or any) writes are processed by the register due to duelling proposers, which makes
it unsuitable to implement basic primitives like counters. The key-value consensus
algorithm Bizur~\cite{DBLP:journals/corr/HochBLV17} is based on a set
of single-writer multi-reader registers and therefore
relies on electing a strong leader.

The use of consistent quorums in conjunction with Paxos is
first introduced by Arad et al.~\cite{consistent-quorum} in the context
of group membership reconfigurations. In this context, a consistent quorum expresses
a consistent \emph{view} of the system in terms of group memberships. Skrzypczak et al.~\cite{DBLP:conf/podc/SkrzypczakSS19}
use consistent quorums to provide linearisable access to CRDTs. While this approach
is similar to the protocol presented here, it heavily relies on the mathematical properties
of CRDTs and can therefore not be used for general state machine replication.

Shared register abstractions were first formalized by
Lamport~\cite{DBLP:journals/dc/Lamport86a}. Among them, the atomic register
provides the strongest guarantees by being linearisable. Numerous implementations
exist today. In particular, the multi-writer generalisation~\cite[p.~25ff.]{DBLP:series/synthesis/2012Vukolic}
of ABD~\cite{DBLP:journals/jacm/AttiyaBD95} has the greatest resemblance to our approach.
However, the properties of atomic registers alone do not suffice to solve consensus,
as not every completed write is necessarily applied to the register when being confronted
with concurrent access. Moreover, only fixed values can be written.
Our register abstractions provide arbitrary value transformations based
on the register's previous value and ensure that completed writes are applied
at-least-once (consensus sequence register) or exactly-once (\approach/).

\section{Conclusion}
In this paper, we introduced register abstractions that satisfy the safety
properties of consensus and allow consensus sequences.
We provided implementations extending
the principles of Paxos consensus, to allow a sequence of consensuses `in-place'
using a single set of storage resources, instead of a separate instance
for every consensus decision.

Additionally, read operations in \approach/
do not interfere with each other (are not serialised with each other) and do not modify any state in the
acceptors when the register is stable, i.e. no write operation is induced. This improves the parallel read throughput and saves
unnecessary, potentially costly state changes of persistent storage
for reads. When reads detect ongoing writes, they can either hope the
writer will finish soon and mitigate the chance of duelling proposers
by just retrying the read, or can start to support the writing
themselves as the writer might have crashed. As we show in our
evaluation (\secref{sec:practical-evaluation}), the trade-off between
both strategies and how often one should retry the read before helping the writer depends on the system
deployment, the number of expected concurrent readers and writers, etc.

Avoiding the need for costly state management and complex protocols
for state pruning, providing fast writing in two message delays and
supporting concurrent readers without serialisation opens a
wide new spectrum of use-cases for Paxos based fault-tolerance. The protocols
we provide are beneficial and applicable on a more fine-grained level than
Multi-Paxos or similar approaches, as they have low system overhead and
provide good scalability.

\medskip
\begin{description}[listparindent=\parindent, topsep=0pt, leftmargin=0cm]
    \item[Code Availability.]
The source code for our \approach/ implementation~\cite{scalaris_approach_impl}
and the protocol scheduler~\cite{scalaris_proto_sched} can be found on GitHub
under the Apache License 2.0.
\end{description}

\section*{Acknowledgements}
We thank Alexander Reinefeld and anonymous reviewers for their dedicated comments and valuable
discussions that helped to improve this manuscript.  This work
received funding from the German Research Foundation (DFG) under grant
RE~1389 as part of the DFG priority program SPP~2037. We thank ZIB's
core facilities unit for providing us the machines and infrastructure
for the evaluation.

\bibliographystyle{IEEEtran}
\bibliography{main.bib}

\vspace{-0.5em}

\begin{IEEEbiography}
    [{\includegraphics[width=1in,height=1.25in,clip,keepaspectratio]{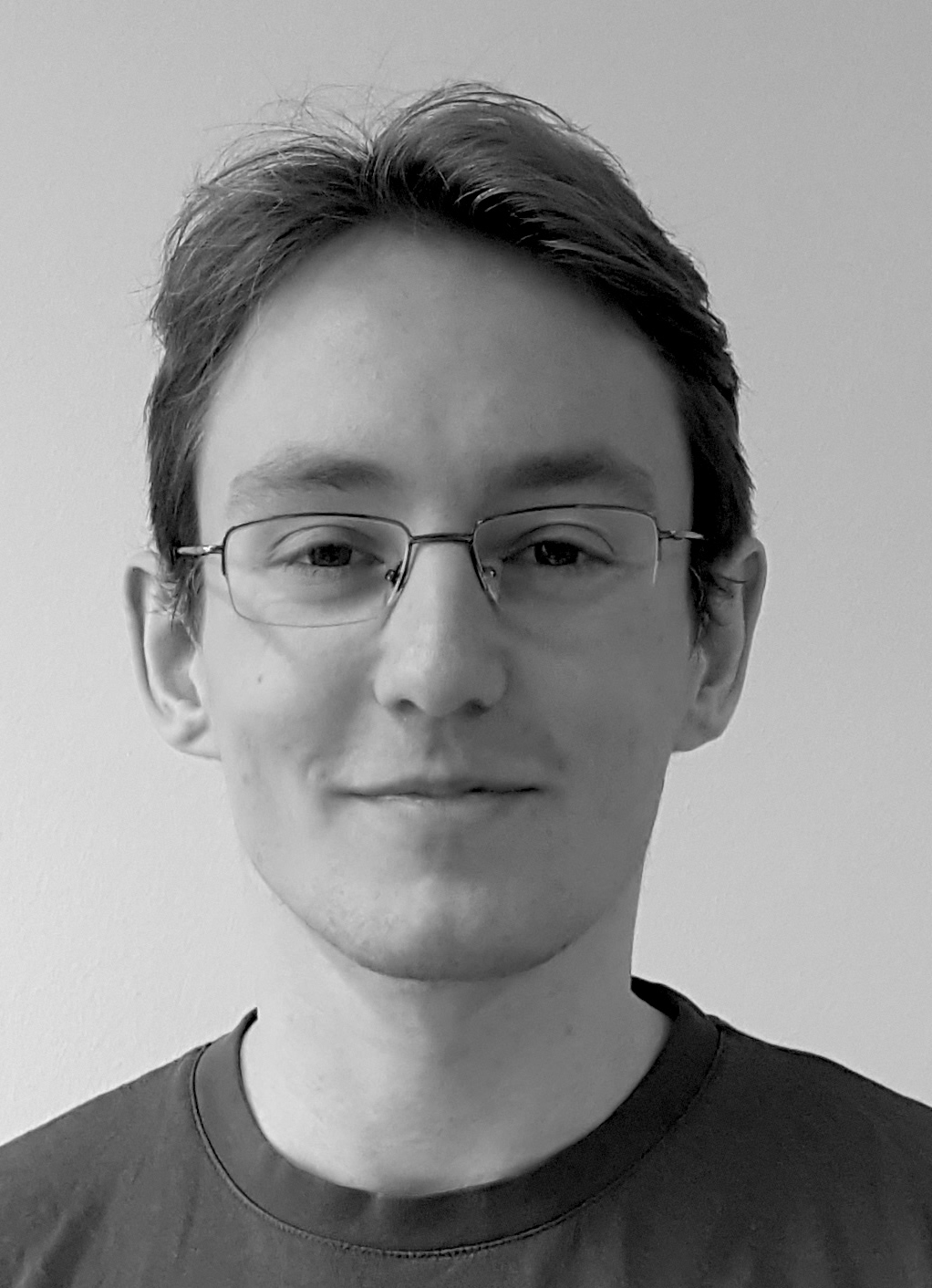}}]
    {Jan~Skrzypczak}
is a research associate at
Zuse Institute Berlin in the department of Distributed Algorithms.
His research interests include the design and implementation of distributed algorithms,
fault-tolerance, reliability and consensus protocols. Skrzypczak received a MSc in
computer science from Humboldt University of Berlin. Contact him at skrzypczak@zib.de.
\end{IEEEbiography}

\begin{IEEEbiography}
    [{\includegraphics[width=1in,height=1.25in,clip,keepaspectratio]{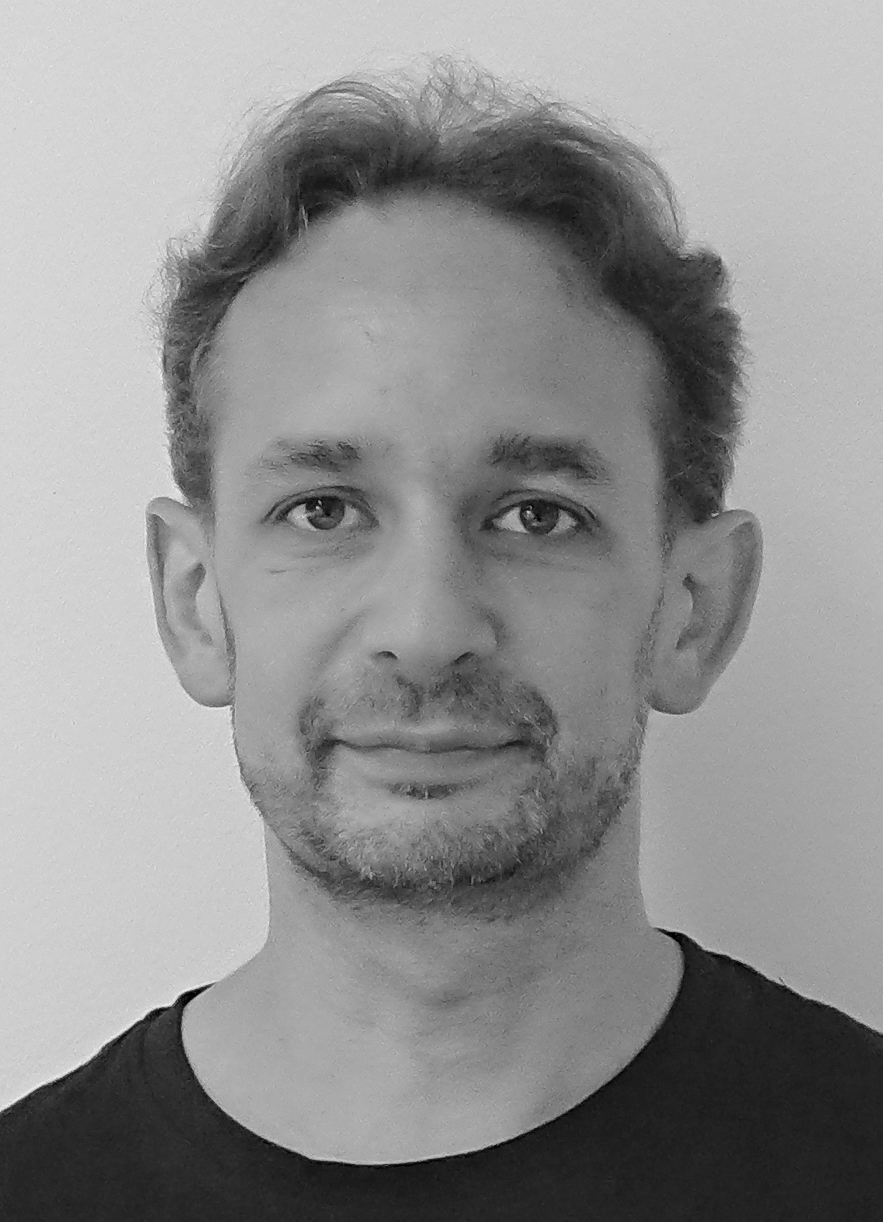}}]
    {Florian~Schintke} is head of the Distributed Algorithms research
    department at Zuse Institute Berlin. His research interests
    include fault-tolerance and scalability, distributed protocols and
    algorithms, transactional key-value stores, and distributed data
    management in general. He received a PhD in computer science from
    the Humboldt-Universität zu Berlin. Contact him at
    schintke@zib.de.
\end{IEEEbiography}

\begin{IEEEbiography}
    [{\includegraphics[width=1in,height=1.25in,clip,keepaspectratio]{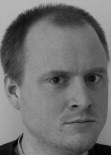}}]
    {Thorsten~Sch{\"u}tt} is a researcher at the Zuse Institute Berlin in the department of Distributed Algorithms. His research interests include P2P protocols, distributed systems, heuristic search, and NVRAM. Sch{\"u}tt received a PhD in computer science from the Humboldt University of Berlin. Contact him at schuett@zib.de.
\end{IEEEbiography}

\end{document}